%% file: ms.tex
\documentclass[twocolumn, switch]{article} %

\usepackage{preprint}

\usepackage{amsmath, amsthm, amssymb, amsfonts}

\usepackage[authoryear]{natbib}
\usepackage[utf8]{inputenc}	%
\usepackage[T1]{fontenc}	%
\usepackage{xcolor}		%
\usepackage[colorlinks = true,
            linkcolor = purple,
            urlcolor  = blue,
            citecolor = cyan,
            anchorcolor = black]{hyperref}	%
\usepackage{booktabs} 		%
\usepackage{nicefrac}		%
\usepackage{microtype}		%
\usepackage{lineno}		%
\usepackage{float}			%
\usepackage{multicol}		%
\usepackage{diagbox}
\usepackage{wasysym}

\usepackage{lipsum}		%

\usepackage{newfloat}
\DeclareFloatingEnvironment[name={Supplementary Figure}]{suppfigure}
\usepackage{sidecap}
\sidecaptionvpos{figure}{c}

\usepackage{titlesec}
\titlespacing\section{0pt}{12pt plus 3pt minus 3pt}{1pt plus 1pt minus 1pt}
\titlespacing\subsection{0pt}{10pt plus 3pt minus 3pt}{1pt plus 1pt minus 1pt}
\titlespacing\subsubsection{0pt}{8pt plus 3pt minus 3pt}{1pt plus 1pt minus 1pt}

\usepackage{tikz,xcolor,hyperref}

\definecolor{lime}{HTML}{A6CE39}
\DeclareRobustCommand{\orcidicon}{
	\begin{tikzpicture}
	\draw[lime, fill=lime] (0,0) 
	circle [radius=0.16] 
	node[white] {{\fontfamily{qag}\selectfont \tiny ID}};
	\draw[white, fill=white] (-0.0625,0.095) 
	circle [radius=0.007];
	\end{tikzpicture}
	\hspace{-2mm}
}
\foreach \x in {A, ..., Z}{\expandafter\xdef\csname orcid\x\endcsname{\noexpand\href{https://orcid.org/\csname orcidauthor\x\endcsname}
			{\noexpand\orcidicon}}
}

\title{Which picker fits my data? A quantitative evaluation of deep learning based seismic pickers}

\usepackage{authblk}

\author[1, 2, *, \#]{Jannes Münchmeyer}
\author[3, \#]{Jack Woollam}
\author[3]{Andreas Rietbrock}
\author[1, 7]{Frederik Tilmann}
\author[6]{Dietrich Lange}
\author[1]{Thomas Bornstein}
\author[4]{Tobias Diehl}
\author[5]{Carlo Giunchi}
\author[4]{Florian Haslinger}
\author[8, 9]{Dario Jozinovi\'c}
\author[8]{Alberto Michelini}
\author[1]{Joachim Saul}
\author[1]{Hugo Soto}

\affil[1]{\footnotesize{Deutsches GeoForschungsZentrum GFZ, Potsdam, Germany}}
\affil[2]{\footnotesize{Institut für Informatik, Humboldt-Universität zu Berlin, Berlin, Germany}}
\affil[3]{\footnotesize{Geophysical Institute (GPI), Karlsruhe Institute of Technology, Karlsruhe, Germany}}
\affil[4]{\footnotesize{Swiss Seismological Service, ETH Zurich, Zurich, Switzerland}}
\affil[5]{\footnotesize{Istituto Nazionale di Geofisica e Vulcanologia, Sezione di Pisa, Pisa, Italy}}
\affil[6]{\footnotesize{GEOMAR Helmholtz Centre for Ocean Research Kiel, Kiel, Germany}}
\affil[7]{\footnotesize{Institut für geologische Wissenschaften, Freie Universität Berlin, Berlin, Germany}}
\affil[8]{\footnotesize{Istituto Nazionale di Geofisica e Vulcanologia, Roma, Italy}}
\affil[9]{\footnotesize{Universit\`a degli Studi Roma Tre, Largo San Leonardo Murialdo 1, Rome, Italy}}
\affil[*]{\footnotesize{corresponding author}}
\affil[$\#$]{\footnotesize{equal contribution}}

\begin{document}

\twocolumn[ %
  \begin{@twocolumnfalse} %
  
\maketitle

\begin{abstract}
Seismic event detection and phase picking are the base of many seismological workflows.
In recent years, several publications demonstrated that deep learning approaches significantly outperform classical approaches and even achieve human-like performance under certain circumstances.
However, as most studies differ in the datasets and exact evaluation tasks studied, it is yet unclear how the different approaches compare to each other.
Furthermore, there are no systematic studies how the models perform in a cross-domain scenario, i.e., when applied to data with different characteristics.
Here, we address these questions by conducting a large-scale benchmark study.
We compare six previously published deep learning models on eight datasets covering local to teleseismic distances and on three tasks: event detection, phase identification and onset time picking.
Furthermore, we compare the results to a classical Baer-Kradolfer picker.
Overall, we observe the best performance for EQTransformer, GPD and PhaseNet, with EQTransformer having a small advantage for teleseismic data.
Furthermore, we conduct a cross-domain study, in which we analyze model performance on datasets they were not trained on.
We show that trained models can be transferred between regions with only mild performance degradation, but not from regional to teleseismic data or vice versa.
As deep learning for detection and picking is a rapidly evolving field, we ensured extensibility of our benchmark by building our code on standardized frameworks and making it openly accessible.
This allows model developers to easily compare new models or evaluate performance on new datasets, beyond those presented here.
Furthermore, we make all trained models available through the SeisBench framework, giving end-users an easy way to apply these models in seismological analysis.
\end{abstract}
\vspace{0.35cm}

  \end{@twocolumnfalse} %
] %

\section{Introduction}

Detecting events and picking seismic phases is at the core of many seismological workflows \citep{bormann2012new}.
It is required for both post-hoc and real-time analysis.
In recent years, several deep learning models for detection and phase picking have been published \citep[e.g.,][]{mousavi2019cred, woollam2019convolutional, ross2018generalized, zhu2019phasenet, soto2020deepphasepick}.
Their excellent performance can largely be attributed to the very large training datasets, with millions of publicly available, manually annotated picks.
A similar abundance of data has led to breakthroughs across domains \citep{lecun2015deep} in the last decade, as deep learning, even among machine learning methods, profits particularly from very large datasets \citep{sun2017revisiting}.

The deep learning based seismic detection and picking methods published so far differ in multiple aspects: their architectures, their training datasets and their task definitions.
These differences currently make it impossible to compare results across publications, in particular as most publications evaluate their model on a single dataset only.
Furthermore, it is often not possible to anticipate how the model will perform on new data that differ from the training data in some characteristics.
Therefore, users seeking to apply deep learning for picking will have difficulties selecting the appropriate model for their task.
Throughout this paper, we will refer to evaluation as "in-domain" when training and test sets come from the same dataset, and as "cross-domain" otherwise.

This study aims to address these issues, by offering a comprehensive benchmark of deep learning methods for detection, picking and phase identification.
In particular, we focus on single station methods, i.e., methods that do not incorporate data from different seismic stations for their picking decision, as not sufficiently many multi-station picking methods have been published yet.
We compare seven models (one classical automatic picking algorithm, six deep learning) on eight datasets to gain a detailed understanding of the models' advantages and disadvantages, when applied for particular tasks or types of data.
We analyzed datasets of different sizes, from different regions, with different arrival-time picking procedures and include a mix of local, regional and teleseismic arrivals.
The deep learning models differ in several points: architecture, with convolutional, recurrent and attention based networks; input representation, in time or frequency domain; output representation, as point or sequence labels; model size, from few layers to very deep models.
We employ consistent training/development/testing splits and parameter selection strategy, to ensure a fair comparison.
We built the benchmark on the SeisBench platform, which we introduce in a companion paper \citep{woollam2021seisbench}, and pytorch lightning \citep{falcon2020lightning}.
Using these frameworks, the benchmark itself is built in a modular way, allowing to add both new datasets and new models to the evaluation easily.
By publishing all code for training and evaluating the models we hope to enable developers of future models to compare their performance to a wide range of known results with minimal effort.

For deep learning pickers there can be a gap between the development of novel methods and their widespread adoption by practitioners.
SeisBench aims to close this gap by offering a unified and simple API, i.e., a standardized programming interface, for applying deep learning models to seismological tasks \citep{woollam2021seisbench}.
As part of SeisBench, we make available the model coefficients for all models trained in the context of this benchmark study. 
Together with our analysis of these models, this enables practitioners to easily select and load the model that is best suited for their specific application scenario.
As this paper is aimed at both machine learning researchers and users with less machine learning expertise, we strive to give a complete description of our evaluation methods, while also providing short explanations for the key ML terms used.

\section{Data and methods}

Multiple deep-learning models for event detection, phase identification and onset picking have been proposed.
However, these models differ with respect to the length of the input waveform and the output specification.
To make the models comparable, we defined three common tasks and define for each model how it is applied to the task.
These tasks are used to evaluate the models.
However, the models might use different data selection and optimization targets in the training phase.
Note that the model training is not tailored to these tasks and is described below.

\textbf{Task 1 - Event detection:} Given a 30~s window of a seismic waveform, determine if it contains an event onset, i.e., a first arrival. We exclude coda examples as it is unclear whether they should be labeled as event or noise.

\textbf{Task 2 - Phase identification:} Given a 10~s window containing exactly one phase arrival, determine if it is a P or an S phase. We do not further differentiate among different P or S phases such as Pn and Pg.

\textbf{Task 3 - Onset time picking:} Given a 10~s window containing exactly one phase arrival of known type (P or S), determine the onset time.

For each of the eight datasets and the three tasks, we generate a set of evaluation targets.
For task 1, we generate noise examples from noise traces, if present in the datasets, or otherwise use windows before the first annotated arrivals in the other traces.
Each evaluation target consists of a three-component waveform window and the associated label.
Models are allowed to use waveforms outside the provided window if they are available in the dataset.

\subsection{Datasets}

We use eight datasets currently included with SeisBench for the benchmark.
Among these datasets, six contain only data from events at local-to-regional distances: ETHZ \citep{woollam2021seisbench}, INSTANCE \citep{michelini2021instance}, Iquique \citep{woollam2021seisbench}, LenDB \citep{magrini2020local}, SCEDC \citep{scedc2013} and STEAD \citep{mousavi2019stanford}.
The other two datasets primarily consist of data from events at teleseismic distances, although including some regional data as well: GEOFON \citep{woollam2021seisbench} and NEIC \citep{yeck2020neic}\footnote{Separate data citations with DOI for the ETHZ and GEOFON datasets are under preparation.}.
The dataset sizes range from 13,400 traces (Iquique) to more than 8 million traces (SCEDC).
For a more detailed description of the datasets, see \citet{woollam2021seisbench}.

All datasets except LenDB contain manually labeled P and S arrivals.
Therefore, we exclude LenDB from tasks 2 and 3, phase identification and arrival time picking.
We note that even though these picks are manually labeled, their exact time is subject to filter selection and human judgement.
Therefore minor discrepancies are to be expected even in case of multiple well-trained human analysts.
Even more detailed phase identification, differentiating, e.g., between Pn and Pg, are available for the ETHZ and GEOFON datasets.
However, within this study, we do not take this fine-grained information into account.
We exclude the NEIC and the Iquique datasets from evaluation for task 1, event detection, as they do not contain either noise examples or sufficiently long waveforms before the arrival to use as noise.
They are, however, used when training the models for evaluation of cross-domain performance.

We resample all datasets to 100~Hz sampling rate if necessary, as this is the original sampling rate used for all models evaluated here.
We note that model performance will be dependent on the sampling rate, but leave this aspect to future study.
To fit the training data and models into 500~GB of main memory, we only train on $90\%$ of the SCEDC training set ($87\%$ for EQTransformer), but use the full development and test set.

\subsection{Models}

\begin{table*}
\begin{center}
\footnotesize
\begin{tabular}{l|p{2cm}p{2cm}p{2cm}p{2cm}p{2cm}p{2cm}}
& BasicPhaseAE & CRED & DPP & EQT  & GPD & PhaseNet\\
\hline
\# Params & 33,687 & 293,569 & 199,731 / 546,081 / 21,181 & 376,935 & 1,741,003 & 23,305\\
Type & U-Net & CNN-RNN & CNN / RNN / RNN & CNN-RNN-Attention & CNN & U-Net\\
Training set & N. Chile & S. California & N. Chile & STEAD & S. California & N. California \\
Orig. weights & N & Y & N & Y & Y & N\\
Reference & \cite{woollam2019convolutional} & \cite{mousavi2019cred} & \cite{soto2020deepphasepick} & \cite{mousavi2020earthquake} & \cite{ross2018generalized} & \cite{zhu2019phasenet}\\
\end{tabular}
\caption{Description of the models studied. The number of parameters refers to the total number of trainable parameters. Note that these numbers might deviate slightly from the ones published by the original authors due to differences in the underlying frameworks. For DPP, information delimited by slashes indicate Detector/P-Picker/S-Picker networks. The row "Orig. weights" indicates whether original weights were published and are available in SeisBench. For PhaseNet, weights were published by the authors, but these weights are not integrated into SeisBench due to technical issues.}
\label{tab:models}
\end{center}
\end{table*}

We evaluate six models for detection and five of these as well for phase identification and onset picking.

BasicPhaseAE \citep{woollam2019convolutional} is a convolutional network for phase detection and onset picking.
It uses a U-Net \citep{ronneberger2015u} like structure.
Input to BasicPhaseAE are 6~s waveforms at 100~Hz and the output are prediction curves for P and S phases and noise with the same length.
BasicPhaseAE was designed to be trained on small datasets and therefore has few parameters to avoid overfitting.
It was originally trained and evaluated on a dataset of 11,000 P/S-pick pairs from the Iquique region in Northern Chile.
For task 1, we use 1 minus the noise probability as the probability of a phase (P or S) being present.
For task 2, we use the ratio of the peak of the P and S predictions.
For task 3, we use the peak position of the relevant phase prediction.

CNN-RNN Earthquake Detector (CRED) \citep{mousavi2019cred} is a pure detection network, that can not be used for phase identification or onset picking.
CRED operates on spectrograms of 30~s waveforms at 100~Hz sampling rate.
Internally, CRED uses convolutional neural network layers (CNN) and long short term memory units (LSTM).
It outputs a prediction curve of 19 samples, indicating whether an earthquake was detected at different times in the signal.
For training, earthquake detection labels are defined based on the P and S arrivals, i.e., detections start at the P arrival and last for 2.4 times the P to S time.
For all datasets without S picks or with teleseismic P arrival, we redefined the detection labels to start at the P arrival and last 20~s.
CRED was originally trained on 550,000 event seismograms and 550,000 noise seismograms from Northern California.
CRED is only tested for task 1, for which we use the peak of the detection.

DeepPhasePick (DPP) \citep{soto2020deepphasepick} is a collection of models for event detection and phase picking.
For detection, it uses a CNN structure with depth-wise separable convolutions, which assigns probabilities for noise, P and S phases to 5~s waveform windows.
Once a P or S arrival is detected, DPP applies the respective picking network.
The picking networks consist of two bidirectional LSTM layers and a pointwise applied fully connected layer.
For picking, the labels are encoded as step functions, with values zero before the onset and one afterwards.
To determine the picking time from the prediction trace, the first prediction sample exceeding 0.5 is used.
The threshold of 0.5 is taken from the original publication.
The three networks, for detection, P picking and S picking are trained separately.
For our study, we use the detection network in tasks 1 and 2.
In task 3, we use the respective pick networks for P and S picks.
We do not use the detection network for task 3, as the window selection for task 3 already gives a good prior on the pick position.
We did not train DPP for S wave picking on GEOFON, due to the very low number of S picks in the dataset.
This issue is not present for the other models, as they are not exclusively trained on one type of arrival.
DPP was originally trained on 25,647 P-phase, 25,647 noise, and 14,397 S-phase windows around the 1995 $M_w=8.1$ Antofagasta and 2007 $M_w=7.7$ Tocopilla earthquakes in Northern Chile.
The original publication of DPP includes an extensive hyperparameter search, i.e., an optimization for the model configuration, with a particular focus on the model architecture.
As our datasets are considerably larger, we are not able to conduct such an optimization here.
Therefore, we chose optimal hyperparameters from the published study, giving us the opportunity to evaluate their transferability to other tasks.

Earthquake transformer (EQTransformer) \citep{mousavi2020earthquake} is a model for joint event detection, phase detection and onset picking.
EQTransformer operates on 60~s waveform windows at 100~Hz sampling rate.
The output of EQTransformer are three prediction traces of 60~s length at 100~Hz sampling rate, each denoting the probability of a detection, P and S wave at a time.
Internally, EQTransformer uses a stack of CNNs, LSTMs and self-attention layers.
In training, EQTransformer makes intensive use of data augmentations.
Here we implemented the same augmentations with the same probabilities $p$: addition of Gaussian noise ($p=0.5$), insertion of gaps ($p=0.2$), dropping of channels ($p=0.3$).
Furthermore, EQTransformer applies a cyclical shift in time to the traces to allow for arbitrary positions of the P and S picks within the window.
We use this augmentation for training on all datasets where not at least 60~s before and after most picks are available, i.e., where the pick cannot naturally occur at any time in the trace.
These datasets are INSTANCE, LenDB, NEIC and STEAD.
We use the same definition for the detection label as for CRED.
EQTransformer was originally trained on STEAD.
For task 1, we use the output of the detection prediction.
For task 2, we use the ratio of the peak of the P and S predictions.
For task 3, we use the peak position of the relevant phase prediction.

Generalized phase detection (GPD) \citep{ross2018generalized} is a phase identification model with a short input window of only 4~s at 100~Hz sampling rate.
For the window, GPD gives one prediction as P, S or noise.
Originally, GPD high-pass filters the input waveforms at 2~Hz.
In contrast, here we use a high-pass filter at 0.5~Hz to take into account that our datasets contain events with lower frequency, in particular in the teleseismic case.
When applying the trained model with a sliding window, the model can also be used for onset detection.
In the original implementations, arrivals were guaranteed to be between seconds 1 and 3 of the input window.
As this can not be guaranteed in our setup, we use a slight modification of the GPD target and loss function used for training the model.
Instead of assigning a class, i.e., noise, P or S, to a window, we assign a probability to each of the classes.
Probabilities for P or S are 1 if the pick is in the center of the window and decline with a Gaussian kernel of width 0.5~s.
To accommodate the modified label definition, we use a multi-class cross-entropy loss for training, similar to the original loss.
For completeness, we provide full results with the original target definition and loss in the supplementary material (Tables S1-S4).
For all tasks, we apply the trained model with a sliding window and a stride of 5 samples, i.e., 0.05~s.
While an even smaller stride might lead to a slight improvement in picking accuracy, it would also come at a considerably higher computational cost, e.g., a stride of 1 would be five times as expensive.
We consider stride 5 as a reasonable balance between accuracy and computational load.
GPD was originally trained and evaluated on 4.5 million seismograms from Southern California with an even distribution between P arrivals, S arrivals and noise.
For task 1, we use 1 minus the noise probability.
For task 2, we use the ratio of the peak of the P and S predictions.
For task 3, we use the peak position of the relevant phase prediction.

PhaseNet \citep{zhu2019phasenet} is a U-Net based model for arrival time picking.
Its inputs are 30~s waveforms at 100~Hz and its outputs are probability curves for P and S arrivals of identical length.
Notably, PhaseNet does not have any "global" connections, i.e., despite its 30~s long input windows, the effective receptive field is only approximately 4~s long.
This means, that predictions at each time are based on relatively small parts of the input data.
From its structure, PhaseNet is fairly similar to BasicPhaseAE, which has been published afterwards.
In contrast to BasicPhaseAE, PhaseNet uses slightly larger filter sizes, has a lower total number of filters, and includes residual connections.
PhaseNet was originally trained and evaluated on 779,514 waveforms with P and S arrivals from Northern California.
For task 1, we use 1 minus the noise probability.
For task 2, we use the ratio of the peak of the P and S predictions.
For task 3, we use the peak position of the relevant phase prediction.
 
\subsection{Training}

We implemented the benchmark using the SeisBench framework \citep{woollam2021seisbench}.
All datasets and models are available in SeisBench and we use augmentations from SeisBench for building training pipelines.
As the length of available waveforms often exceeds the expected input lengths of the models, we selected windows according to the following schema.
In 2/3 of the cases, we selected a window such that at least one pick is guaranteed to be within the window.
In the remaining cases, we randomly select a window from the full trace, which can also contain picks.
This strategy ensures that the training labels are not dominated by noise examples, in particular for models with short input windows.
We apply the same window selection for generating training and development examples.
Note that this window selection strategy is not to be confused with the window selection for the three evaluation tasks.
The window selection here selects windows of appropriate length for training the models.
The windows selected for the tasks are identical across all models and their length is independent of the specific model.

We did not conduct any resampling between P and S arrivals, as the number of available P and S picks are always within a factor of 4 of each other, i.e., no massive label imbalance is present.
Only for the GEOFON dataset, the label imbalance is strong, with 100 times more P than S arrivals.
However, as only around 2,800 S arrivals are available for this dataset, we found the number insufficient for effective upsampling.
This label imbalance for GEOFON will be taken into account during evaluation.
For the DPP pickers, we only train on examples containing either P or S picks, as the picker assumes that exactly one pick is within the window.

We train the models using the Adam optimizer \citep{kingma2014adam}.
We trained each model for 100 epochs, but with an additional limit of 48 hours wall time.
This wall time limit only terminated training of some models on the very large SCEDC dataset, but the validation loss curves strongly suggested that the models had been fully trained already nonetheless.
In total, training and evaluation of the models, including cross-domain evaluations, took $\sim4000$ GPU hours and $\sim260,000$ CPU thread hours.
A breakdown of the computational costs for the different models is contained in the discussion.

\subsection{Threshold and hyperparameter selection}
\label{sec:parameter_selection}

We train all models on the training parts of the datasets.
For evaluation, we use the model with the lowest loss, i.e., the metric scoring the quality of the models predictions in training, on the development set.
For task 1 we evaluate the area under the receiver operating characteristic (ROC-AUC or short AUC), which is independent of the decision threshold between noise and event.
However, we also show optimal configurations in terms of F1 score for reference.
The F1 score is defined as the harmonic mean of precision, the fraction of correct detections among all detections, and recall, the fraction of detections among all events.
It is therefore a combined measure for both the sensitivity and the specificity of a model.
For task 2 we choose the decision threshold between P and S phase to optimize the Matthews correlation coefficient (MCC).
MCC is a symmetric metric with values between -1 (total disagreement) and 1 (full agreement) and is regarded as a well suited measure for binary classification performance even in case of class imbalance.
For each of the tasks we select the optimal thresholds on the development sets.
We select thresholds independently for each model and dataset.
The thresholds are documented in the supplement (Tables S5-S6).
For cross-dataset analysis, we select the model based on the loss on the development set of the source dataset and select the threshold on the development set of the target dataset.
If not indicated otherwise, all results reported are from the test parts of the datasets.

For all models, we used a fixed batch size of 1024 samples.
We ran all experiments with learning rates, i.e., the step size for the gradient descent optimization algorithm, of $10^{-2}$, $10^{-3}$ and $10^{-4}$.
Learning rates were kept constant during the full training.
We selected the best performing model based on the development set of the target using F1 score (task 1), MCC (task 2) or standard deviation (task 3), both for in-domain and cross-domain analysis.
Due to the huge computational demand, we were not able to conduct a large scale hyperparameter study.
Nevertheless, we are confident the test results are reliable, as similar hyperparameters were used in the original publications and the Adam optimizer is known to require only low levels of hyperparameter tuning \citep{kingma2014adam}.

\subsection{Baseline}

For P onset time picking we include a traditional picker, the Baer-Kradolfer picker \citep{baer1987automatic}, as baseline.
The Baer-Kradolfer picker depends on four parameters: a minimum required time to declare an event, a maximum time allowed below a threshold for event detection, and two thresholds.
For details on the parameters, we refer to \citet{baer1987automatic} or \citet{kueperkoch2012}.
We set the second threshold to half of the first threshold to reduce the number of parameters.
Furthermore, the Baer-Kradolfer picker expects a bandpass filtered signal, therefore we add two additional parameters to be tuned, the high- and low-pass frequencies of a causal Butterworth-bandpass-filter.

In contrast to the deep learning models, the parameters for the Baer-Kradolfer picker can not be optimized using gradient descent.
Therefore, we optimize parameters using Gaussian optimization with the root mean squared error (RMSE) as fitness function.
We use 25 initial points and 500 further evaluations of the fitness function.
To reduce computational demand, we only evaluate the fitness on 2500 P picks from the development set.
We use the same 2500 P picks for each evaluation of the fitness function.
This does not severely limit model performance, as the number of parameters is very low, with only five parameters to select.

We do not include classical baselines for either detection or S wave picking.
For detection, the classical workflow includes a picker with rather high false positive rate, followed by event association.
As our datasets and experimental setup do not allow to run the association step, this approach could not be employed here.
Furthermore, association will likely also improve the performance of the deep learning pickers.
We do not include a classical S picking baseline, as they usually require additional event information, e.g., the approximate event-station distance or the back-azimuth, and careful manual tuning.
Classical S pickers often have considerably more parameters than classical P pickers that need to be adjusted.
For example, the picker presented by \citet{diehl2009automatic} has 14 parameters (see \citet[Table 4]{diehl2009automatic}), not including the parameters for frequency filtering or quality classification.
Tuning these parameters is not feasible with simple optimization, but requires informed judgement for each individual dataset.

\section{Results}

\subsection{Task 1 - Event detection}

\begin{figure*}
    \centering
    \includegraphics[width=0.6\textwidth]{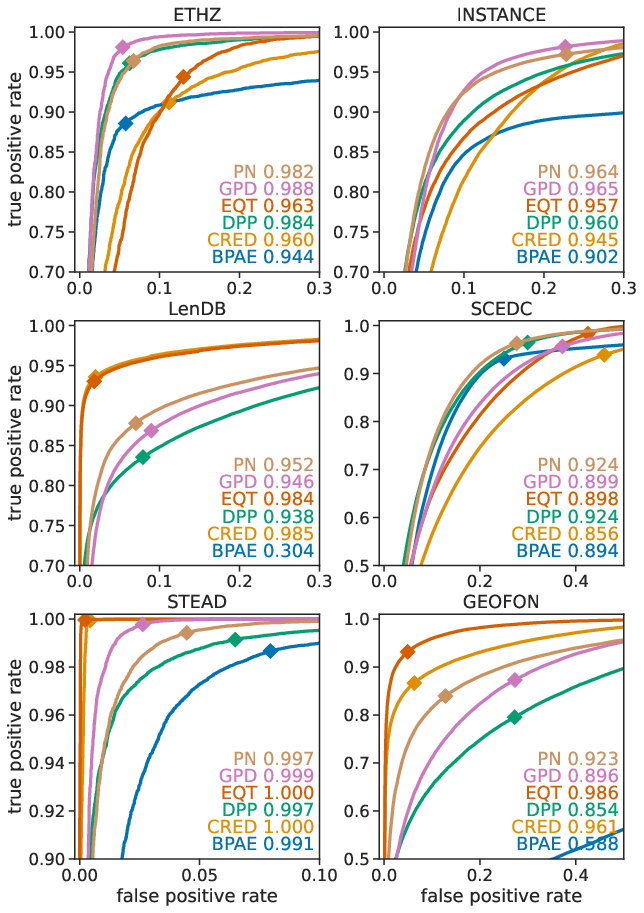}
    \caption{Receiver operating characteristic for detection results from in-domain experiments. Each panel shows one dataset, each curve one model. Models were selected to maximize AUC score. Numbers in the corners indicate the test AUC scores. Markers indicate the point with the configuration associated with the highest F1 score. If no marker is shown, the optimal configuration is outside the shown range. Note that axis ranges differ between the subplots. A similar plot showing the full curves is available in the supplementary material (Figure S1).}
    \label{fig:detection_roc}
\end{figure*}

We evaluate the first task, event detection, using receiver operating characteristics (ROC) and the corresponding area under the curve (AUC).
An AUC of 1 indicates a perfect model, an AUC of 0.5 a coin toss.
We use the ROC because, depending on the application scenario, different trade-offs between false positive rate and true positive rate are required.
For example, when using a simple pick association algorithm in a region with low seismic activity, a false positive rate below 0.01 might be required, while when using a hyperbolic pick association \citep{woollam2020hex} in a seismically active region, a false positive rate of 0.05 might be absolutely fine.
This tuning can be achieved by using different decision thresholds.
To complement the ROC with a single number that can easily be compared, we use the AUC, which gives an average performance across all possible thresholds.
We note that AUC values can be dominated by the asymptotic behavior of the ROC curves, however, in our results the AUC values clearly correspond to the visual conclusions from the ROC and an inspection of the full curves showed no unexpected asymptotic behavior.

Results from in-domain analysis are available in Figure \ref{fig:detection_roc}.
On average, EQTransformer shows the best performance (AUC 0.964), closely followed by PhaseNet (0.957), CRED (0.951), GPD (0.949) and DPP (0.943).
Further behind is BasicPhaseAE (0.771).
The considerably worse performance of BasicPhaseAE compared to PhaseNet is surprising, given their very similar architecture.
However, this shows that the shorter input windows for BasicPhaseAE, together with the shorter filters and missing residual connections lead to considerably worse results.
Overall, CRED and EQTransformer show similar performance to each other for all datasets, and also GPD and PhaseNet show similar performance to each other.
This can be explained with the similar architectures: EQTransformer is an extended version of CRED, using attention structures in addition to the CNN and RNN structures.
Similarly, the architectures of GPD and PhaseNet both use CNNs on a relatively short input window, and in that sense PhaseNet can be interpreted as a GPD-like network with sequence instead of point predictions.

While EQTransformer and CRED on average perform better than GPD, PhaseNet and DPP, this results exclusively from better performance on LenDB and GEOFON.
In both cases, we argue that the longer receptive fields of these models allow for the better performance.
For GEOFON, the lower frequency signals of teleseismic arrivals can likely be better captured with these longer receptive fields, therefore representing a genuine improvement.
For DPP, which performs particularly badly on GEOFON, the reason might also be that its hyperparameters were explicitly tuned on a local seismic dataset, i.e., a dataset with very different characteristics.
Even though the other models were also built for local or regional data, their hyperparameters were not tuned in a similarly systematic fashion as for DPP.
In contrast to the GEOFON case, for LenDB, the global view of the input window gives CRED and EQTransformer the ability to learn the characteristics of the dataset;  the first arrival is always at a similar location within the 27~s input window, which gives them an (unfair) advantage over the other models.
In addition, GPD, PhaseNet and DPP suffer from the inaccurate pick times in LenDB, which result from using predicted arrival times for labelling.
The characteristic training function for detection thus can start either too late or too early, meaning it is hard to minimise loss globally in training. 

On INSTANCE, SCEDC and STEAD no systematic differences between the models except BasicPhaseAE can be observed.
However, on ETHZ GPD, PhaseNet and DPP outperform CRED and EQTransformer by a small margin of $\sim$0.02 points AUC score.
The difference likely results from the definition of detections in the different models and the types of picks in the ETHZ dataset.
GPD, PhaseNet and DPP simply calculate their detection score as one minus the noise probability.
In contrast, CRED and EQTransformer provide explicit detection curves, that are fitted to predefined detection labels.
Following the original publications, this detection label depends on P and S position.
Therefore, at least for the regional datasets, detections were only declared if both P and S waves were annotated in the dataset.
In the ETHZ dataset for a considerable number of traces either P or S annotations are missing, thus negatively affecting the performance of EQTransformer and CRED.

\subsection{Task 2 - Phase identification}

\begin{table*}
 \centering
 \small
 \caption{Phase identification results from in-domain experiments given by Matthews correlation coefficient (MCC). Averages are macro-averages, i.e., the same weight is given to all models and datasets.}
\input{tables/phase_test_mcc.tex}
\label{tab:phase_test_mcc}
\end{table*}

We evaluate task 2 using the Matthews correlation coefficient (MCC).
The MCC is symmetric, i.e., in contrast to the AUC or F1 score independent on a choice of positive and negative class.
It takes values between -1 (total disagreement) and 1 (full agreement).
In-domain results for all datasets and models are available in Table \ref{tab:phase_test_mcc}.
For phase identification, EQTransformer shows the best results with an average MCC of 0.95, followed by GPD (0.90), PhaseNet (0.86), DPP (0.76) and BasicPhaseAE (0.71).
These differences are considerably larger than for detection, indicating that the ability of EQTransformer to incorporate waveforms from a larger time window to understand the context indeed improves phase identification performance.
In terms of data sets, phase identification is similarly hard for the five regional datasets (average MCC 0.90), more difficult for NEIC (0.81) and even more difficult for GEOFON (0.50).
Again, the worse performance on the teleseismic dataset most likely results from the lower frequency content of the arrivals.
In addition, GEOFON only contains $<3000$ S picks in total, leading to a very small training set for these.

\subsection{Task 3 - Onset time determination}

\begin{figure*}
    \centering
    \includegraphics[width=\textwidth]{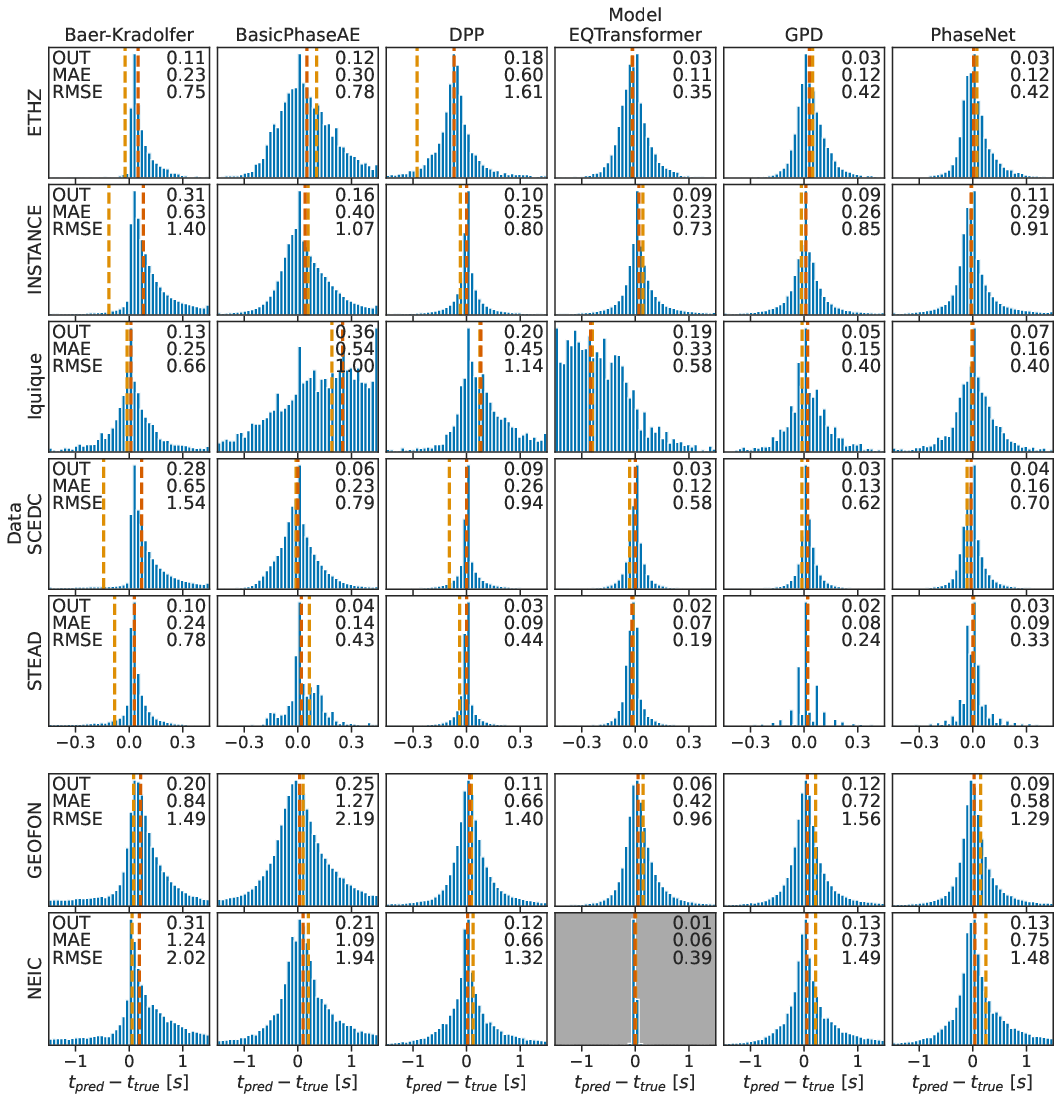}
    \caption{Histogram of P residuals from in-domain experiments. The numbers in the corner indicate the fraction of samples outside the plot boundaries, the mean absolute error (MAE) and the root mean squared error (RMSE). Vertical dashed lines show median (red) and mean (orange) of the residuals. For enhanced visibility, y axis scaling differs between all panels, therefore bar heights can not be compared across panels. Note also the different x axis scales for regional and teleseismic datasets. The EQTransformer on NEIC panel shows invalid results due to data constraints. For computation of the MAE, RMSE, mean and median of the Baer-Kradolfer picker we exclude picks within the first second of the window as these are mostly invalid. Due to the underlying obspy implementation, traces where no pick can be generated are picked early in the trace, leading to these picks. These picks are included for calculating the fraction of samples outside the plot boundaries.}
    \label{fig:test_P_diff}
\end{figure*}

\begin{figure*}
    \centering
    \includegraphics[width=\textwidth]{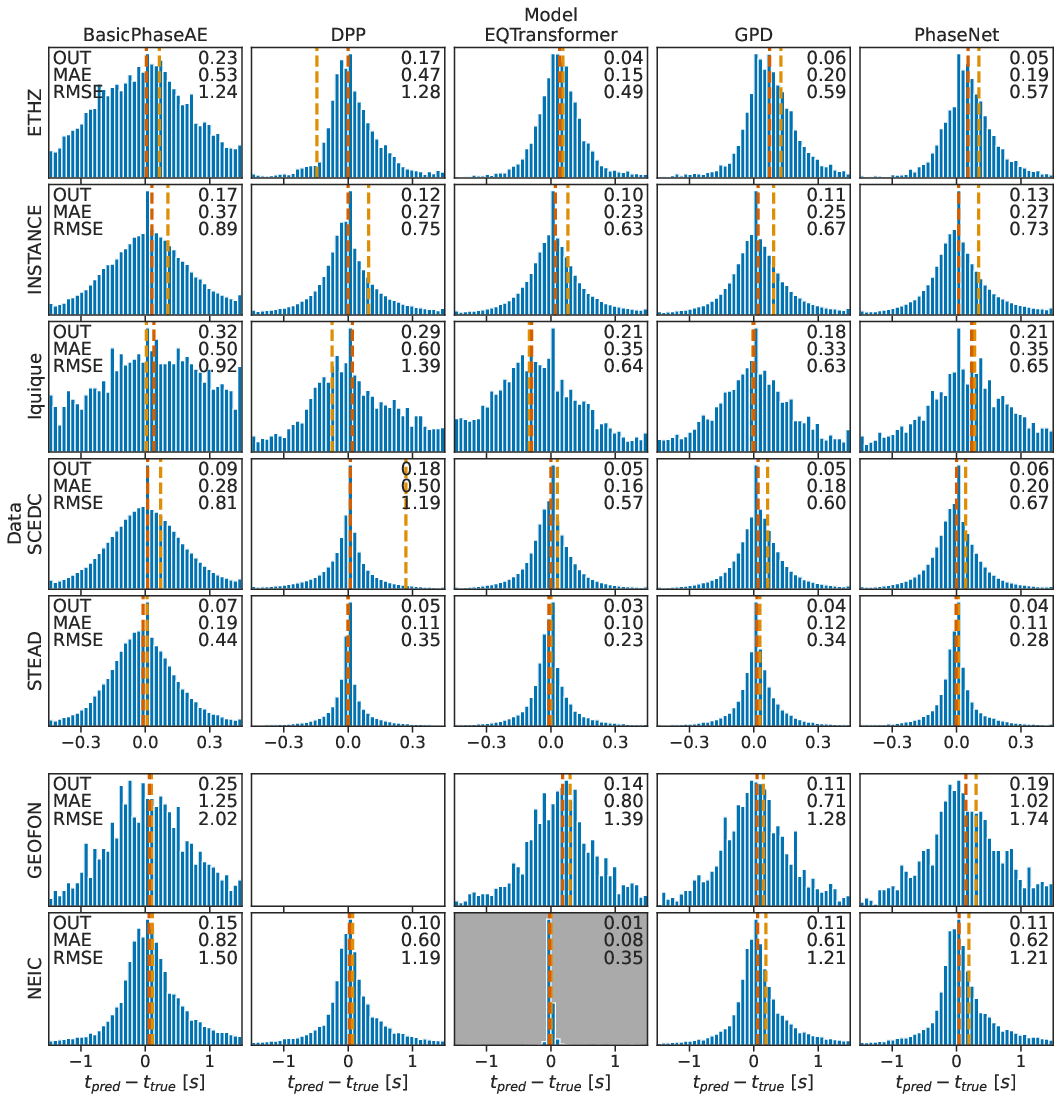}
    \caption{Histogram of S residuals from in-domain experiments. The numbers in the corner indicate the fraction of samples outside the plot boundaries (OUT), the mean absolute error (MAE) and the root mean squared error (RMSE). Vertical dashed lines show median (red) and mean (orange) of the residuals. For enhanced visibility, y axis scaling differs between all panels, therefore bar heights can not be compared across panels. Note also the different x axis scales for regional and teleseismic datasets. The EQTransformer on NEIC panel shows invalid results due to data constraints. Results for DPP on GEOFON are not available due to insufficient training data.}
    \label{fig:test_S_diff}
\end{figure*}

For evaluating task 3, we use the residuals, i.e., the differences between ML-pick time and the manual reference pick time.
We analyze the fraction of samples with high residuals ($>0.45$~s for regional, $>1.5$~s for teleseismic), the root mean squared error (RMSE), and the mean absolute error (MAE).
We evaluate both RMSE and MAE, to include one metric that is sensitive to outliers (RMSE) and one that is not (MAE).
In-domain P arrival picking results are shown in Figure \ref{fig:test_P_diff}, results for S arrival picking in Figure \ref{fig:test_S_diff}.
On almost all datasets, for both P and S waves, EQTransformer performs best, although usually only with a small margin to GPD, PhaseNet and in some instances DPP.
We note that the exceptional performance of EQTransformer on the NEIC dataset results from an artifact.
In the NEIC traces, the picks are always at the same position within the 60~s input window.
As EQTransformer has a global view of the full 60~s window, it does not need to learn to actually identify pick positions, but only to output a constant position.
As this is not realistic for an actual application, and the other models can not reproduce this artifact due to their short input windows, this result needs to be excluded from the interpretation.

For both P and S waves, BasicPhaseAE performs worst on most datasets, with both RMSE and MAE often more than twice those of the best model.
For P waves, DPP performs similarly to EQTransformer, GPD and Phasenet on STEAD, GEOFON, INSTANCE and NEIC, but considerably worse on ETHZ, Iquique and SCEDC.
DPP results on S waves mirror the P results, with competitive performance on INSTANCE, STEAD and NEIC, but considerably worse performance than the best models on ETHZ, Iquique and SCEDC.
From our observations, this behavior is likely caused by the unstable training of the LSTM in the DPP picker.
The sequence length of 1000 samples is fairly long for an LSTM and can lead to vanishing gradients.
We observed that validation losses for DPP showed very high fluctuation over the training duration, with some training runs for some learning rates even failing to converge at all.
As this is a random effect, it might be possible to improve performance to some extent by retraining.

Among EQTransformer, GPD and PhaseNet, on the five regional seismic datasets, performance for both P and S waves is usually similar.
On these datasets, for P waves, EQTransformer consistently shows 0.01 to 0.04 lower MAE than GPD and PhaseNet.
For S waves, absolute performance differences are slightly larger, likely due to the higher absolute errors, but again EQTransformer shows the best performance on the regional datasets.
As an exception, EQTransformer shows considerably worse performance on the Iquique dataset, in particular for P waves.
Given that small size of the Iquique dataset (13,400 examples), we think that the number of training examples might be insufficient for the complex EQTransformer architecture.

On the teleseismic GEOFON dataset, EQTransformer has considerably lower MAE for P waves (0.42~s) than GPD (0.72~s) and PhaseNet (0.58~s).
Similar to task 1, this can likely be explained with the longer receptive field of EQTransformer being beneficial for the lower frequency content of teleseismic signals.
However, this effect can not be observed for the S picks in GEOFON.
Here, GPD performs best in terms of MAE (0.71~s), followed by EQTransformer (0.80~s) and PhaseNet (1.02~s).
Due to the small number of S picks in the GEOFON dataset (<3,000), these results are, however, not representative for the performance on teleseismic data in general, but are rather related to the performance in a low training data scenario.

Comparing the average performance of the models on the different datasets, taking into account only the three best models, there are consistent differences.
For P waves, the lowest average MAE values occur for STEAD (0.08~s), ETHZ (0.12~s), SCEDC (0.14~s), Iquique (0.21~s) and INSTANCE (0.26~s).
The higher MAE for INSTANCE might partially result from the relatively large group of human annotators ($\sim$15--20 people) that contributed to the dataset, leading to slightly lower consistency in the picks.
Considerably higher MAE values were determined for GEOFON (0.57~s) and NEIC (0.74~s, excluding EQTransformer).
These higher residuals can be explained with the teleseismic traces.
For these, the onset times are often more challenging to pick due to their emergent onsets and lower frequency contents compared to mostly impulsive regional arrivals.
Furthermore, the teleseismic arrivals in the GEOFON and NEIC datasets often exhibit worse signal to noise ratios.

For the S waves, average MAEs are approximately $25 \%$ to $60 \%$ worse than the respective P residuals.
We observe two exceptions with differing behavior.
First, for INSTANCE, average S residuals are even slightly lower than the corresponding P residuals, leading to similar S residuals as for ETHZ or SCEDC.
We think that this might be caused by a higher quality of the S picks compared to the P picks in INSTANCE.
(for an example of an inaccurate P label, see Fig~\ref{fig:predictions}a).
Second, for NEIC, average S residuals are even considerably lower than the P residuals.
We speculate that this is an artifact of the dataset creation: in teleseismic analysis, S waves are picked less regularly than P waves, which is also reflected in the lower number of S waves in the dataset.
However, this also means that S waves tend to be picked primarily in more favorable signal to noise conditions and at shorter distances, both of which lead to better defined pick onsets and in turn to lower residuals for the models.

We now analyze the residual histograms for P (Figure \ref{fig:test_P_diff}) and S arrivals (Figure \ref{fig:test_S_diff}).
With the exception of the artifact for EQTransformer on NEIC, all residual distribution roughly resemble Laplacian densities with different widths, i.e., distributions with a sharp mode and relatively heavy tails.
Nearly all distributions are exactly centered on zero, i.e., their mode is at zero.
In some cases, the modes seem to be more centered than the mean error, indicating outliers to be systematically biased towards either too early or too late estimation.
On all datasets, the residual distributions show no systematic differences between the best performing models, i.e., no model exhibits, for example, a particular skew.

In contrast to the deep learning pickers, the classical Baer-Kradolfer picker shows considerably different features.
First, the residual distribution is clearly non-symmetric, with considerably higher likelihood of the model picking slightly late than slightly early.
This is expected, as the Baer-Kradolfer picker can not pick before the energetic onset.
Second, while the Baer-Kradolfer picker shows similarly low residuals as the best deep learning models for the majority of picks, it has a considerably higher fraction of outliers.
Therefore, in favorable signal-to-noise conditions, a well tuned Baer-Kradolfer picker is competitive with the deep learning models; in less favorable conditions, the deep learning pickers perform considerably better.

\subsection{Cross-domain performance}

\begin{figure*}
    \centering
    \includegraphics[width=0.9\textwidth]{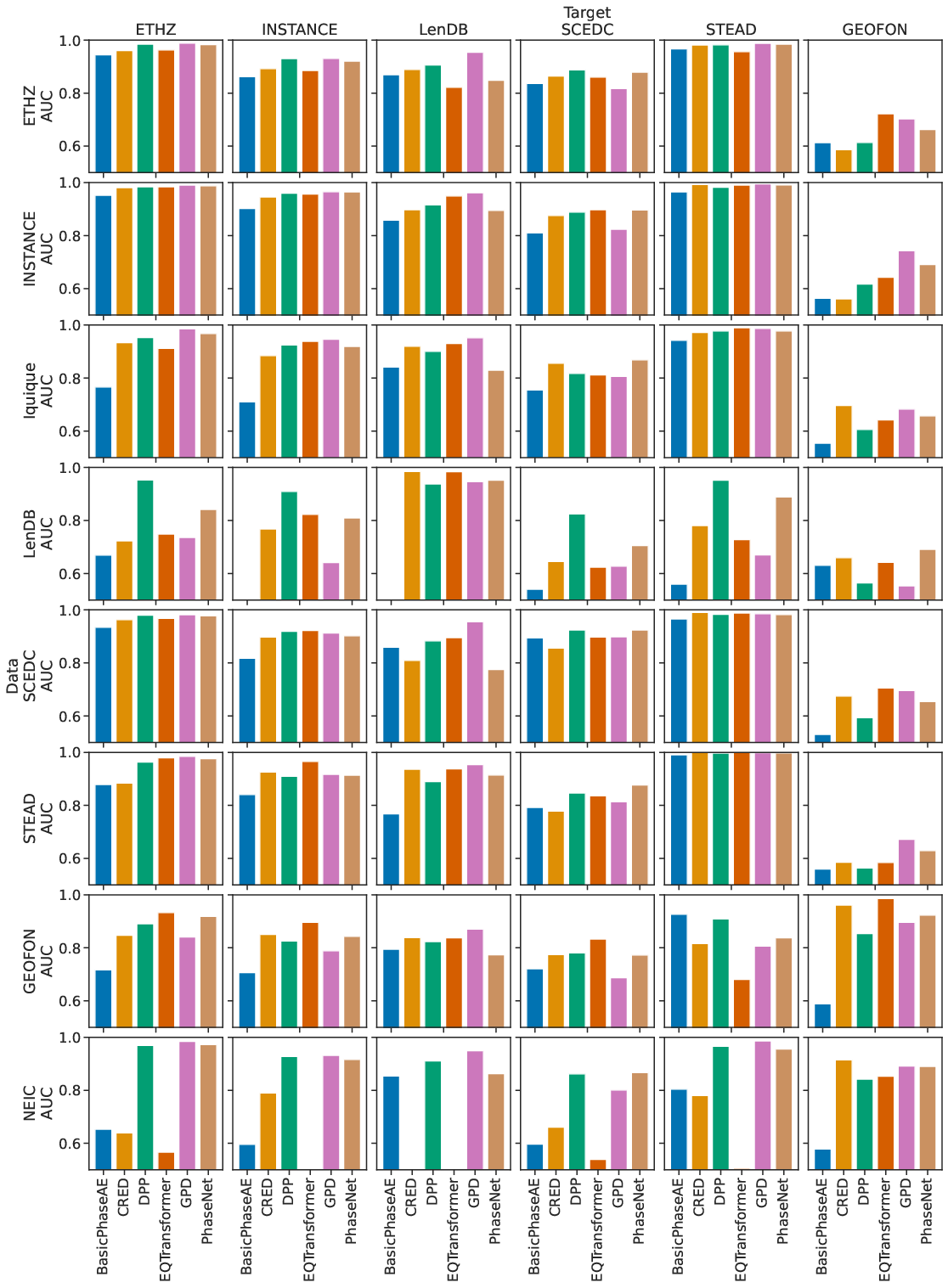}
    \caption{AUC scores for detection results from cross-domain experiments. Each panel shows one combination of training (row) and evaluation (column) dataset, each bar one model. Models were selected to maximize AUC score on the evaluation dataset. Note that the bars start at 0.5 instead of 0 as 0.5 is the AUC of a random model. Bars not shown have AUC values below 0.5. A figure with the ROC curves and the numerical AUC values can be found in the supplement (Figure S2).}
    \label{fig:detection_roc_cross}
\end{figure*}

\begin{figure*}
    \centering
    \includegraphics[width=\textwidth]{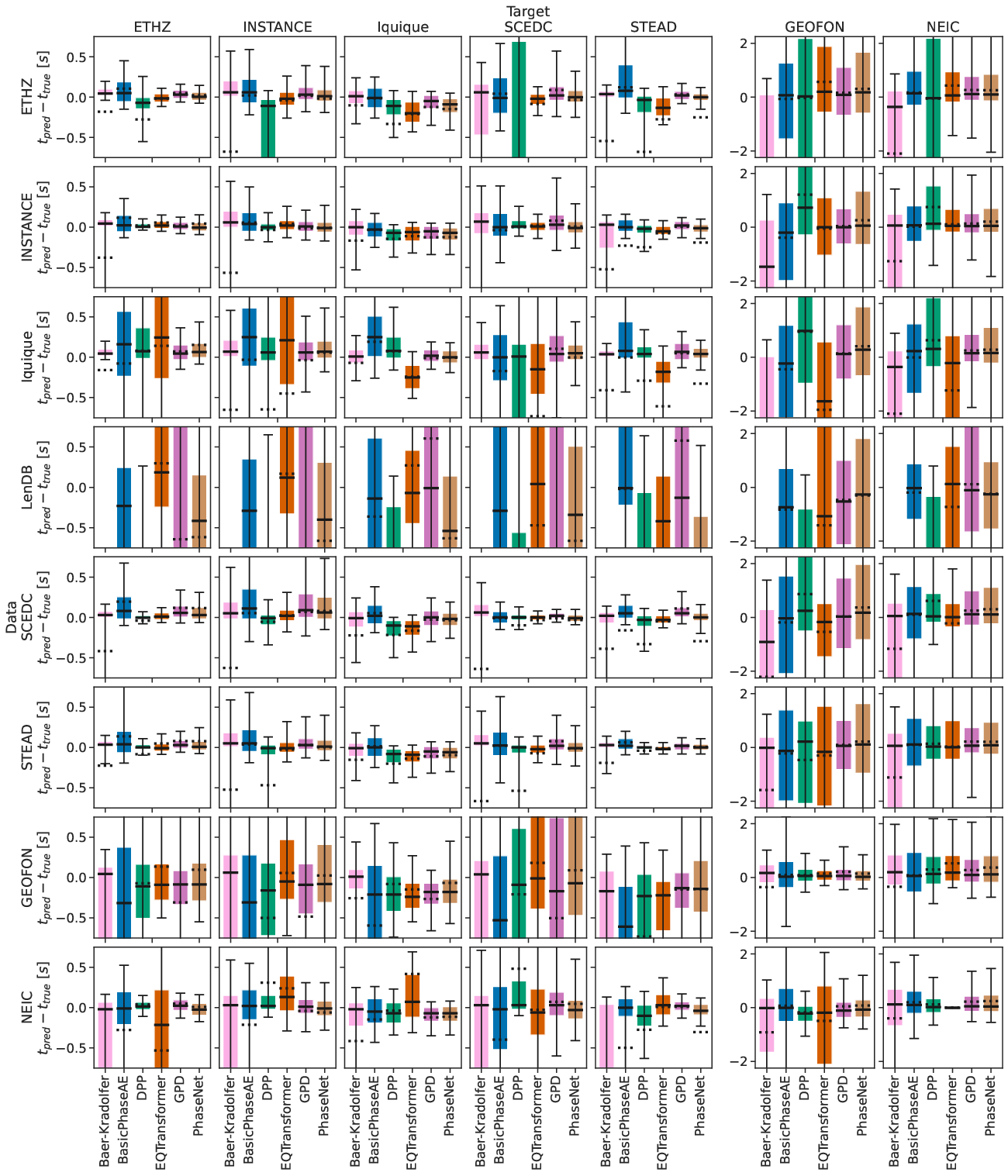}
    \caption{Distribution of P pick residuals from cross-domain experiments. Each panel shows one combination of training (row) and evaluation (column) dataset, each bar one model. The solid bars show the interquartile range, the whiskers range from the 10th to the 90th percentile. The solid lines indicate medians, the dashed lines indicate means. Note that we include LenDB as a training dataset, but not as an evaluation dataset, because learning to pick on the predicted arrival times might be possible, while they do not serve as a sufficient reference for evaluating picking performance. An analogous plot for S pick residuals is available in the supplementary material (Figure S3).}
    \label{fig:p_diff_cross}
\end{figure*}

So far, all presented results were in-domain results, i.e., the models were trained and tested on datasets with mostly identical characteristics.
However, in practice, one will often need to deviate from this principle and apply a model trained on one dataset to different data.
The performance of models in this cross-domain setup can be considerably different from the in-domain performance, because the characteristics of the target data might be different from those of the training data, but also because particularities in manual reference picking or selection might lead to unexpected biases in the trained models.
To evaluate how the models fare in a cross-domain application, we perform a cross evaluation of the models, i.e., we take each trained model and evaluate it on all test datasets on which it was not trained.
Due to the vast number of results ($\#$ models $\times$ $\#$ source datasets $\times$ $\#$ target datasets = 336) for each task, we only report selected results in the main text (Figure \ref{fig:detection_roc_cross}, Figure \ref{fig:p_diff_cross}).
The full results for all tasks are available in the supplementary material (Tables S7-S12, Figure S2-S16).

The most obvious result from the cross-domain study is that, in general, cross-application works well if both datasets contain traces from the same distance range, although usually worse than in-domain application, but completely fails when applying regional models to teleseismic data.
For example in task 1, no model trained on a regional dataset reaches an AUC score above 0.75 for detection on GEOFON, which is considerably below the in-domain performance of 0.85 to 0.99.
When trained on NEIC, CRED (F1 0.86), EQTransformer (F1 0.78), GPD (F1 0.82), and PhaseNet (F1 0.82) achieve good F1 scores on GEOFON, mostly comparable with the in-domain performance.
The opposite case, applying teleseismic models to regional data, also shows considerably worse performance than between regional datasets, but performance degradation is not as bad as from regional to teleseismic.
However, this might at least be partially caused by the fact that both the GEOFON and NEIC dataset contain at least some examples of regional picks.
Notably, for determining pick onset times, regional to teleseismic application and vice versa work, at least to some extent, with residual distributions being wide, but clearly centered around 0.

The models trained on two of the regional datasets, Iquique and LenDB, show worse cross-domain performance than the ones trained on the other datasets.
For Iquique, we suspect that this results from the small training set, leading to less well defined model parameters and lower generalization ability.
For LenDB, this most likely results from the picks being obtained from travel-time calculation with a velocity model instead of manually labeled phase arrival times and the short input windows, leading to a data bias in the learned models.
Notably, the DPP model strongly outperforms the other models for detection, when trained on LenDB.
This is most likely caused by the label definition for detection with DPP: it only considers if a P/S pick is contained, but does not incorporate its position, making it less affected by the inaccurate pick positions in the dataset.
A data bias is also visible for EQTransformer trained on NEIC.
As mentioned above, picks in the NEIC dataset are always at the same position in the 60~s waveform traces, which can be recognized by EQTransformer.
Applications of EQTransformer trained with NEIC data therefore perform considerably worse than other models with the same combination.
However, the performance is still significantly better than would be expected in case of a constant pick location, indicating that the data augmentation employed in the training of EQTransformer indeed enables it to partially mitigate this issue.

For detection between different regional datasets (excluding LenDB and Iquique), PhaseNet performs best (AUC 0.947), closely followed by EQTransformer (0.941), GPD (0.937) and DPP (0.934).
CRED shows similar performance to EQTransformer in most cases, but shows considerably worse performance in a few cases, in particular when trained on STEAD.
For phase identification in the same setup, EQTransformer works best (MCC 0.95), outperforming PhaseNet (0.82), and GPD (0.76).

When evaluating picking performance, a clear feature is that all models incur an elevated level of picks with large differences ($>$0.45~s/1.5~s) in cross-domain application compared to in-domain application.
The fraction of picks with such large differences often goes up to 10\% even for datasets with generally good results in cross-application, e.g., STEAD and INSTANCE.
We suspect that this might be caused by differences in annotation practice, i.e., some datasets might tend to miss weak earlier phase arrivals or might not exclude examples with overlapping events.
For P arrival picking at regional distances, in most cases EQTransformer performs best.
However, this result is not fully consistent, with several cases of GPD outperforming EQTransformer, sometimes even considerably, e.g., from ETHZ or INSTANCE to STEAD.
PhaseNet performs slightly worse than EQTransformer in most cases, but for some combinations works even considerably worse, e.g., from SCEDC to ETHZ.
However, we are not able to identify a systematic pattern when a specific model shows particularly good or bad cross-domain performance among the regional datasets.

In some cases, the cross-domain application also reveals biases in the data.
For example, the P picks from all models trained on NEIC and applied to GEOFON are systematically too early.
Conversely, trained on GEOFON and applied to NEIC, the arrivals are picked too late.
This indicates that the two agencies exercise different judgement when picking arrivals.

On average, models trained on INSTANCE perform best in all tasks.
However, differences for detection performance are usually minor with models achieving similarly good detection results when trained on STEAD, SCEDC and to some extent ETHZ, as when trained on INSTANCE.
For P wave picking, performance of models trained on INSTANCE is usually better than on other datasets, even though training on STEAD also yields good performance.
We note that this is in contrast to the in-domain performance, where models consistently showed worse performance on INSTANCE than on STEAD.
Possibly the better performance for models trained on INSTANCE can be explained with the higher average number of waveforms per event (21 for INSTANCE, 2 for STEAD), leading to more diverse waveforms for each event.
Models trained on ETHZ and SCEDC often show worse detection performance, but perform nearly on par with models trained on INSTANCE for picking.
The overall good performance of INSTANCE and STEAD can likely be explained with a combination of the quality, the size and the diversity of the datasets.
Both datasets contain more than one million P picks and more than 700,000 S picks, giving plenty of training examples for the models.
While SCEDC contains even more picks, the higher diversity in the picks in the other datasets likely leads to the better performance of models trained on INSTANCE and STEAD.
For STEAD, this diversity is achieved by including picks from different regions.
For INSTANCE, the diversity results from the complex tectonic setting of Italy, giving rise to both crustal seismicity and subduction events.
These results indicate that for training a transferable model, it is highly desirable to include diverse picks.

\section{Discussion}

\subsection{Model comparison on waveform examples}

\begin{figure*}
    \centering
    \includegraphics[width=\textwidth]{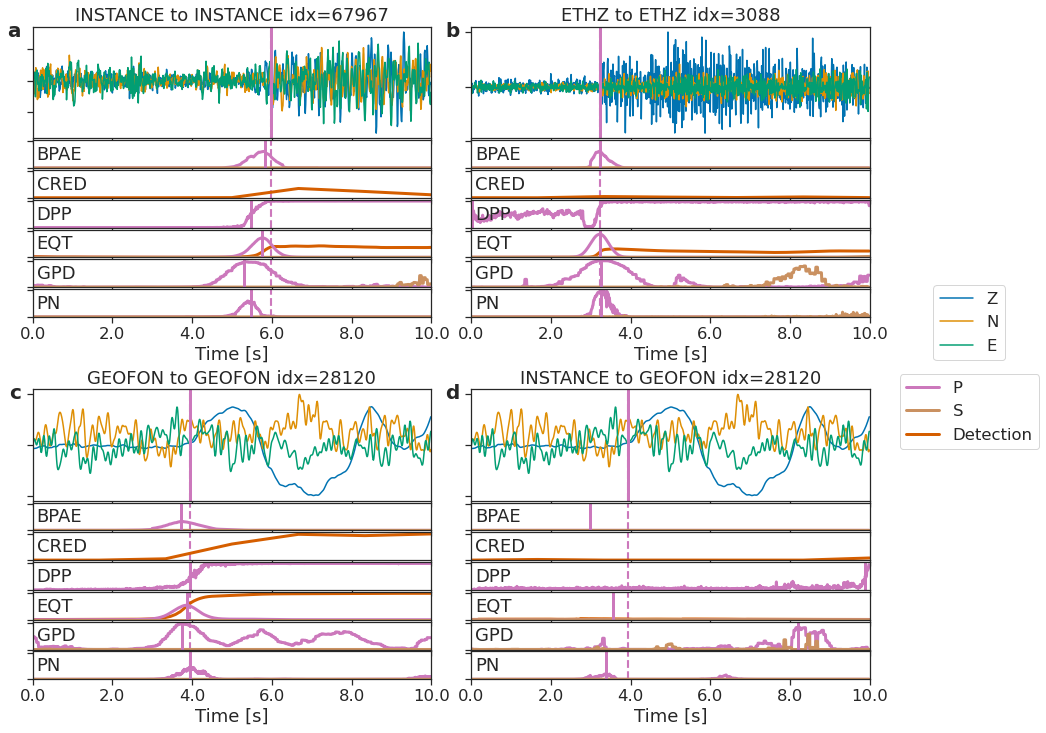}
    \caption{Example predictions for waveforms from different dataset. \textbf{a} INSTANCE ($\sim50$~km epicentral distance, $M_L=2.0$). Waveforms were highpass filtered at 2~Hz for better visibility of the onset time. \textbf{b} ETHZ ($\sim65$~km, $M_L=2.2$). \textbf{c} GEOFON ($\sim7,300$~km, $M_w=5.7$). \textbf{d} Same waveforms as in c (GEOFON), but annotated with models trained on the INSTANCE dataset. In all plots, vertical lines indicate pick times. The manual pick times are indicated by solid vertical lines in the waveforms plot and dashed lines in the prediction plots. Note that the ML-predicted pick for DPP in panel b is at 0.0~s.}
    \label{fig:predictions}
\end{figure*}

For further insights into the models, we present several examples for which we compare the predictions from the different models (Figure \ref{fig:predictions}, a-c in-domain, d cross-domain).
All examples are from the test sets.
Figure \ref{fig:predictions}a shows predictions around a P pick from the INSTANCE dataset.
In this example, all models pick 0.5~s before the annotated onset.
Indeed, when inspecting the waveforms after a highpass filter at 2~Hz, the pick seems to be annotated roughly 0.5~s too late in the original dataset.
This highlights the ever possible imperfections in the training datasets deriving from oversights of the analysts as in this case.
On the other hand, it also illustrates how the deep learning models are able to learn a more consistent picking than present in the dataset, because they can not reproduce differences between human annotators.

Figure \ref{fig:predictions}b shows predictions around a P pick from the ETHZ dataset.
BasicPhaseAE, EQTransformer, GPD and PhaseNet all correctly identify the P pick and determine the onset time to within 0.1~s.
DPP fails to correctly identify the onset time.
While the prediction curve correctly jumps from 0 to 1 around the pick, it already exceeds 0.5 within the first seconds, leading to an early pick.
Predictions curves are smooth for EQTransformer, while predictions from the other models are considerably more rough.
This presumably results from the long-range relationships modeled in EQTransformer but not accounted for in the other models.
Except for GPD, no model detects additional potential picks within the trace.
However, we observed that GPD operates best when choosing a very high detection threshold, such that the secondary picks would be ignored in practice.

Figures \ref{fig:predictions}c and \ref{fig:predictions}d both show the same teleseismic P arrival from the GEOFON dataset.
However, while Figure \ref{fig:predictions}c shows the predictions for models also trained on GEOFON, Figure \ref{fig:predictions}d shows predictions for models trained on INSTANCE, i.e., without teleseismic arrivals in the training data.
When trained on the GEOFON data, all models correctly detect the pick and its time with an error below 0.2~s.
Notably, even in this scenario GPD detects multiple secondary P picks, indicating difficulties in differentiating the onset of the low frequency signal ($\sim0.25$~Hz) from its later wiggles.
In contrast to the models trained on GEOFON, the models trained on INSTANCE consistently miss the arrival and return mostly arbitrary onset times.
Only GPD produces detections, but as mentioned above, GPD detections should only be treated as true picks for very high confidence scores, which are not reached in this example.
This example confirms the conclusion from the quantitative cross-domain analysis that cross-domain application only works well within the same distance range.

\subsection{Cross-domain application with adjusted sampling rate}

As reported above, models trained on regional data perform poorly when applied to teleseismic examples.
As a key reason, we identified the lower frequency content of the teleseismic arrivals.
To further validate this hypothesis and to test a mitigation strategy, we analyzed rescaled versions of trained models.
To this end, we apply the models trained on waveforms with 100~Hz sampling rate to waveforms with a considerably lower sampling rate.
As the models do not know about timing, but only relative sample position, this effectively downscales the frequency ranges the models are looking for.
We applied models trained on ETHZ, INSTANCE, Iquique, SCEDC and STEAD to the GEOFON and NEIC datasets.
We tried sampling rates of 20~Hz and 40~Hz, i.e., downscaling by factors of 5 and 2.5.
For each combination of model, target and source dataset, we selected the combination of learning rate and target sampling rate that showed best development scores and report the test results.
As before, we only report selected results in the main text, but full results are available in the supplement (Tables S13-S15, Figures S17-S20).

For event detection (task 1) on GEOFON, resampling improves performance considerably, with AUC values up to 0.908 (GPD trained on INSTANCE).
This is even above the in-domain score of GPD on GEOFON.
While the non-resampled models achieve AUC scores often only slightly above a trivial classifier, the resampled models consistently outperform the trivial classifier.
This also indicates that it might be reasonable to directly train the models on 20 or 40~Hz teleseismic data to achieve better performance.
We did not conduct this test here, but leave it for future study.
However, we note that CRED (0.961) and EQTransformer (0.986) still achieve better in-domain results.
Still, this is close to the optimal cross-domain performance on GEOFON (0.915), achieved with CRED trained on NEIC.
As NEIC is not applicable to task 1, we can only report the GEOFON results.

Similarly to detection, for P wave picking (task 3) we observe substantial improvements for EQTransformer, GPD and PhaseNet.
Best performance on GEOFON is achieved with PhaseNet trained on INSTANCE (MAE 1.01~s).
For NEIC, the same model achieves an MAE of 0.90~s.
For GEOFON, this is considerably inferior to the best model trained on NEIC without resampling (MAE 0.77~s) and the best in-domain performance (MAE 0.66~s).
For NEIC, this score is superior to the best cross-domain model (MAE 0.95~s), but again does not outperform the optimal in-domain model (MAE 0.73~s).
The error distributions are similar to the original error distributions, in particular they are centered around zero in most cases and they have heavy tails.
The fraction with large residuals $>1.5$~s exceeds 20~\% or even 30~\% in most cases.

In contrast to the P wave case, we do not see an improvement for S wave onset determination with the resampled models.
As reported above, the original models already performed considerably better for S wave detection than for P wave detection in both in- and cross-domain analysis.
We explained this with the selection procedure, leading mostly to S picks with good signal to noise ratio and at moderate distances.
For the same reason, we expect the typical amplitude spectra of regional and teleseismic datasets to be more similar for S arrivals than for P arrivals.
Therefore, resampling does not yield performance improvements for S waves.

\subsection{Computational demand}

A major consideration for deep learning models is their computational demand.
We trained all models on identical machines, always using one Nvidia A100 GPU with 40 GB GPU memory.
Except for the LSTM of the DPP picker and for evaluating GPD with the sliding window approach, we never got close to using the full memory, with all models staying well below 10 GB at a batch size of 1024 samples.
We note that larger batch sizes could have lead to better performance, however, we decided not to experiment with larger batch sizes as they were not employed in the original publications.
Furthermore, large batch sizes often require the use of specific optimizers, e.g., LARS \citep{you2017large}, which tend to be less robust than the Adam optimizer.

To quantify the performance of the models, we measured run times for training and evaluation on INSTANCE.
We chose INSTANCE for two reasons: it is large enough to ensure run times are not dominated by the overhead of epoch starts and ends, and it naturally comes at a sampling rate of 100~Hz and therefore does not require resampling on the fly, which could lead to CPU saturation.
In our measurements, we did not include overheads from data preloading and model setup.
We focus our analysis on throughput in training and evaluation.
However, we note that for training performance, this only gives a rough guidance, as convergence speeds might differ between the models.
For incorporating this aspect, models need to be compared provided a fixed compute budget.
As for most applications the inference time is of bigger concern than the training time, we do not conduct this analysis here.

For training, the fastest models were BasicPhaseAE and the DPP detection network, both with $\sim7500$~samples per second.
The other models achieved, in decreasing order, PhaseNet ($\sim6300$), GPD ($\sim5700$), CRED ($\sim4900$), DPP picker networks ($\sim3100$), and EQTransformer ($\sim2600$).
For evaluation, again BasicPhaseAE achieved the highest throughput with $\sim6700$~sample per second.
The other models achieved, in decreasing order, CRED ($\sim4900$), DPP detection network ($\sim4800$), PhaseNet ($\sim4800$), DPP picker networks ($\sim4700$), EQTransformer ($\sim3000$) and GPD ($\sim64$).
The very poor throughput of GPD in evaluation results from its sliding window approach.
While all other models give prediction curves, GPD only gives point predictions and therefore needs to be applied repeatedly for each trace at regular intervals.
We chose a stride of 5 samples, i.e., applied GPD every 5 samples, as a good balance between accuracy and runtime.
However, GPD is still slower than the next slowest model by a factor $\sim50$.
Overall, performance differences are within a factor of $\sim2$, except for the evaluation of GPD.
However, we note that performance differences might be considerably different on other hardware, in particular systems without GPUs and older hardware.

While the provided numbers give an indication of the model performance on our hardware, they do not immediately imply which resource is limiting the performance, i.e., if the models are CPU bound, GPU bound or if a memory bus saturates.
From observations of computing resources utilization during training and evaluation, we are confident that EQTransformer, the DPP pickers and CRED are GPU limited.
The same holds true for GPD in evaluation.
For the remaining models, we experienced GPU loads considerably below $100\%$, indicating a limitation on CPU or memory bus side.

\section{Conclusions \& Recommendations}

In this study, we conducted a quantitative comparison of six deep learning based models for earthquake detection, phase identification and onset time determination.
Using eight datasets - six of local and regional distance recordings and two mainly at teleseismic distances - we evaluated both in- and cross-domain performance.
In conclusion, we found EQTransformer, GPD and PhaseNet to be the best performing models.
Among these three models, EQTransformer shows considerably better performance for teleseismic data, likely due to its longer receptive field.
GPD, while showing excellent performance, only achieves poor throughput in evaluation, making it only applicable for small datasets or with large computational resources being available.
PhaseNet achieves similar performance to GPD, while providing significantly higher throughput.
CRED and DPP also achieve very good detection performance.
However, CRED, in contrast to the other models, is limited to detection, which makes it less appealing, in particular considering its close architectural similarity to EQTransformer.
DPP is performing well on detection but shows considerably poorer performance for onset time determination.

The results of our study do not only represent a model comparison, but also give guidance which training datasets should be used in a cross-domain application, e.g. %
when the pretrained weights provided with this study through SeisBench are used on new data.
The most important factor is using a training dataset from the appropriate distance range.
For local to regional data, models trained on STEAD and INSTANCE generally showed best performance.
Combined with the model discussion above, we would recommend using PhaseNet or EQTransformer for picking and detection on these datasets.
For teleseismic targets, models trained on teleseismic datasets should be used.
As our comparison only used two teleseismic dataset, we can not give a clear recommendation here.
However, caution needs to be exercised when using EQTransformer trained on NEIC due to the fixed pick positions.
If only detections, but no pick locations are required, CRED trained on the datasets mentioned above is a viable alternative.

While our study analyzed both in- and cross-domain performance, it exclusively focused on event based analysis, i.e., we only analyzed the models' performance on pre-selected windows.
We chose this approach, as it is a good first-order proxy for the model performance in practical applications and as it allows for a thorough quantitative evaluation.
Furthermore, it is closely related to a practical application scenario: post-processing.
For example, NEIC applies deep learning models to the outputs of their STA/LTA pickers, to refine pick times and estimate phase type and event-station distance.

A different use case would be applying the models to continuous data, e.g., for creating seismicity catalogs.
While our results still give some guidance for this case, the different properties of the problem need to be taken into account.
For example, in a continuous setup the false positive rate needs to be significantly lower than in post-processing, as only a fraction of all windows will usually contain arrivals, leading to a strongly biased prior distribution.
Furthermore, assumptions used in the benchmark might become incorrect, e.g., windows might contain multiple picks, in particular in dense aftershock sequences.
A detailed analysis of the performance of the models applied to continuous data therefore should be conducted in a follow up study.

Another application case for deep learning pickers is real-time identification of earthquake arrivals.
The results from this study only considered the post hoc performance of the models, not taking into account how early they would be able to identify an event onset.
This aspect needs to be studied explicitly, before applying the models in a real-time/early warning scenario.

Besides the performance evaluation presented in this paper, our study also yielded a rich collection of trained models.
We make trained model weights for all combinations of datasets and models publicly available through the SeisBench framework.
These models can be used directly by practitioners wanting to automatically pick their data, but they can also be used for further evaluation as discussed above.
For the appropriate choice of models, this study should give a good guideline.
For each model, we also provide suggested decision thresholds through SeisBench.
Even though the optimal threshold will depend on the application scenario, the provided values give an orientation for threshold selection.
In addition, these models can serve as a basis for transfer learning, which recently has been shown to considerably improve the performance of deep learning in a seismological context \citep{munchmeyer2021transformer, jozinovic2021transfer}.
We expect this to be particularly beneficial when training on catalogs of limited size (<10,000 events).

\section*{Data and Resources}

SeisBench is available at \url{https://github.com/seisbench/seisbench} and \url{https://doi.org/10.5281/zenodo.5568813}. The benchmarking code is available at \url{https://github.com/seisbench/pick-benchmark}. All data used can be accessed through SeisBench.

\section*{Acknowledgements}

This work was supported by the Helmholtz Association Initiative and Networking Fund on the HAICORE@KIT partition.
JM acknowledges the support of the Helmholtz Einstein International Berlin Research School in Data Science (HEIBRiDS).
We thank the Impuls- und Vernetzungsfonds of the HGF to support the REPORT-DL project under the grant agreement ZT-I-PF-5-53.
This work was also partially supported by the project INGV Pianeta Dinamico 2021 Tema 8 SOME (CUP D53J1900017001) funded by Italian Ministry of University and Research ``Fondo finalizzato al rilancio degli investimenti delle amministrazioni centrali dello Stato e allo sviluppo del Paese, legge 145/2018''.

\normalsize
\bibliography{bibliography}

\clearpage
\appendix

\renewcommand\thefigure{S\arabic{figure}}
\renewcommand\thetable{S\arabic{table}}
\renewcommand\thesection{S\arabic{section}}
\setcounter{figure}{0} 
\setcounter{table}{0}

\section{Supplementary material}

\begin{table*}
 \centering
 \small
 \caption{Detection results including the original GPD variant}
\input{tables/gpd/detection_test_gpd.tex}
\label{tab:detection_test}
\end{table*}

\begin{table*}
 \centering
 \small
 \caption{Phase identification results including the original GPD variant}
\input{tables/gpd/phase_test_mcc_gpd.tex}
\label{tab:phase_test}
\end{table*}

\begin{table*}
 \centering
 \small
 \caption{Accuracy of P picks including the original GPD variant}
\input{tables/gpd/precision_p_test_gpd.tex}
\label{tab:precision_p_test}
\end{table*}

\begin{table*}
 \centering
 \small
 \caption{Accuracy of S picks including the original GPD variant}
\input{tables/gpd/precision_s_test_gpd.tex}
\label{tab:precision_s_test}
\end{table*}

\begin{table*}
 \centering
 \small
 \caption{F1 optimal thresholds for event detection}
\input{tables/detection_thresholds.tex}
\label{tab:detection_thresholds}
\end{table*}

\begin{table*}
 \centering
 \small
 \caption{MCC optimal thresholds for phase identification}
\input{tables/phase_thresholds.tex}
\label{tab:prhase_thresholds}
\end{table*}

\begin{figure*}
    \centering
    \includegraphics[width=0.6\textwidth]{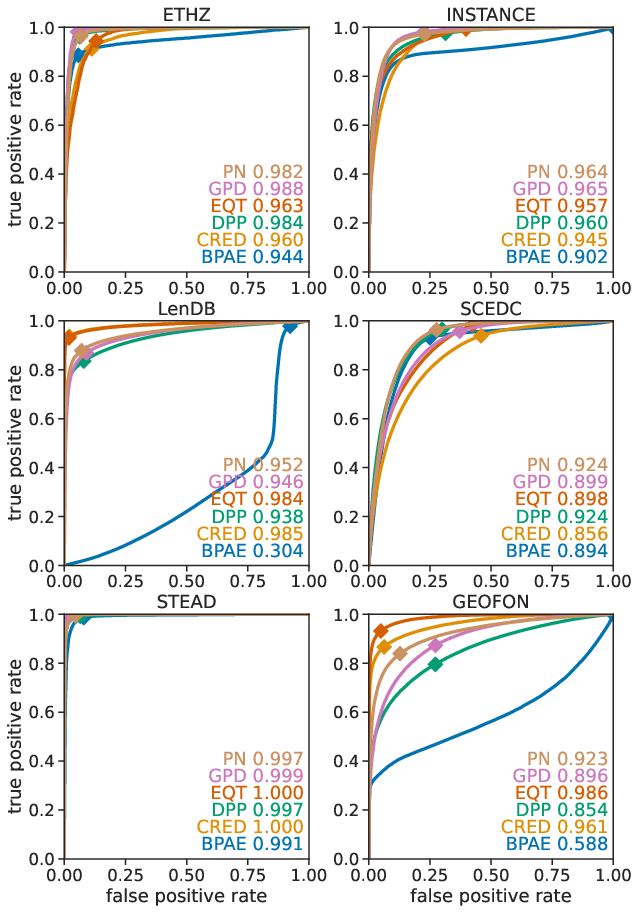}
    \caption{Receiver operating characteristic for detection results from in-domain experiments. Each panel shows one dataset, each curve on model. Models were selected to maximize AUC score. Numbers in the corners indicate the test AUC scores. Markers indicate the point with configuration with highest F1 score.}
    \label{fig:detection_roc}
\end{figure*}

\begin{figure*}
    \centering
    \includegraphics[width=0.9\textwidth]{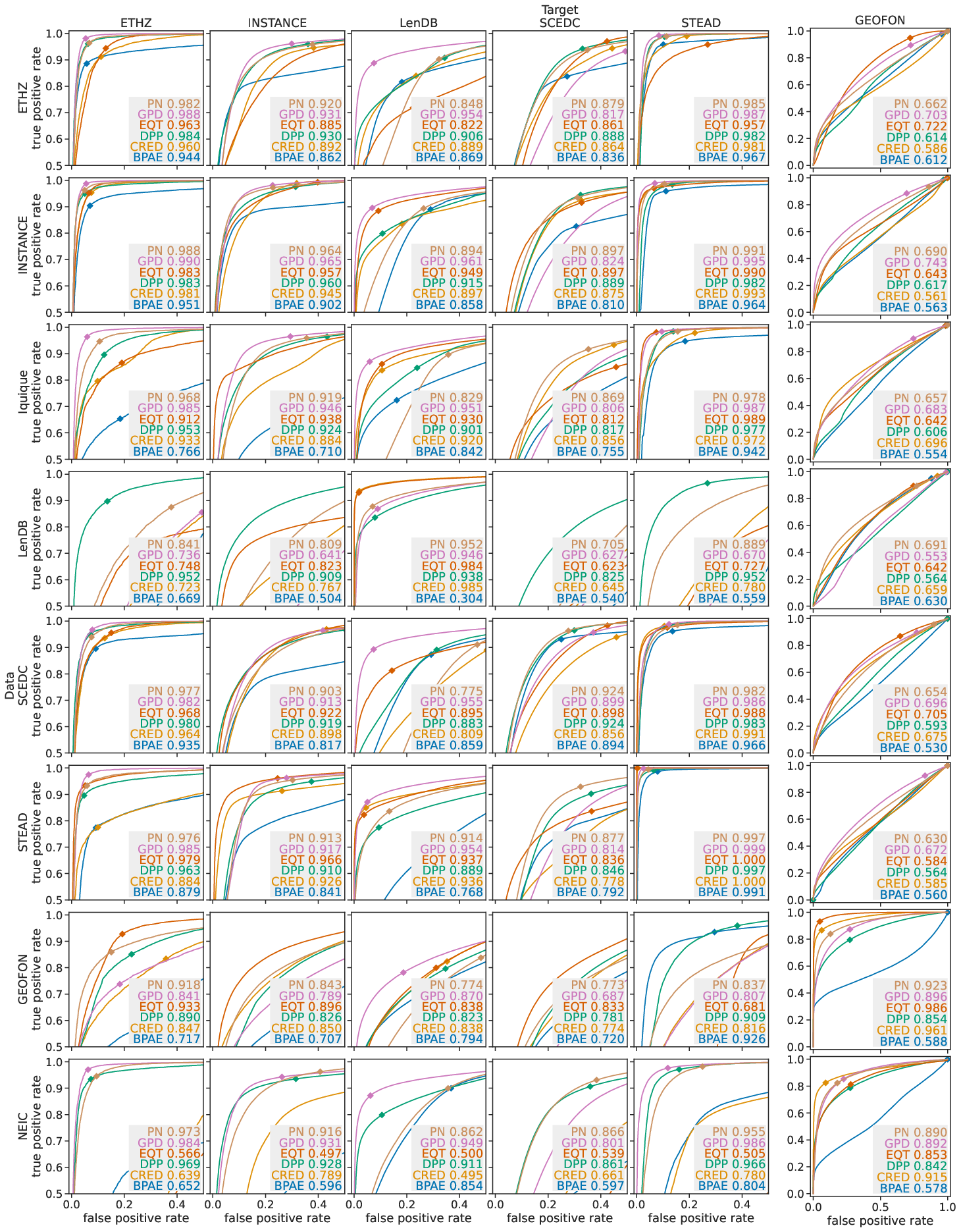}
    \caption{Receiver operating characteristic for detection results from cross-domain experiments. Each panel shows one combination of training (row) and evaluation (column) dataset, each curve one model. Models were selected to maximize AUC score on the evaluation dataset. Numbers in the corners indicate the test AUC scores. Markers indicate the point with configuration with highest F1 score. If no marker is shown, the optimal configuration is outside the shown range. Note that axis ranges differ between the subplots.}
    \label{fig:detection_roc_cross}
\end{figure*}

\begin{figure*}
    \centering
    \includegraphics[width=\textwidth]{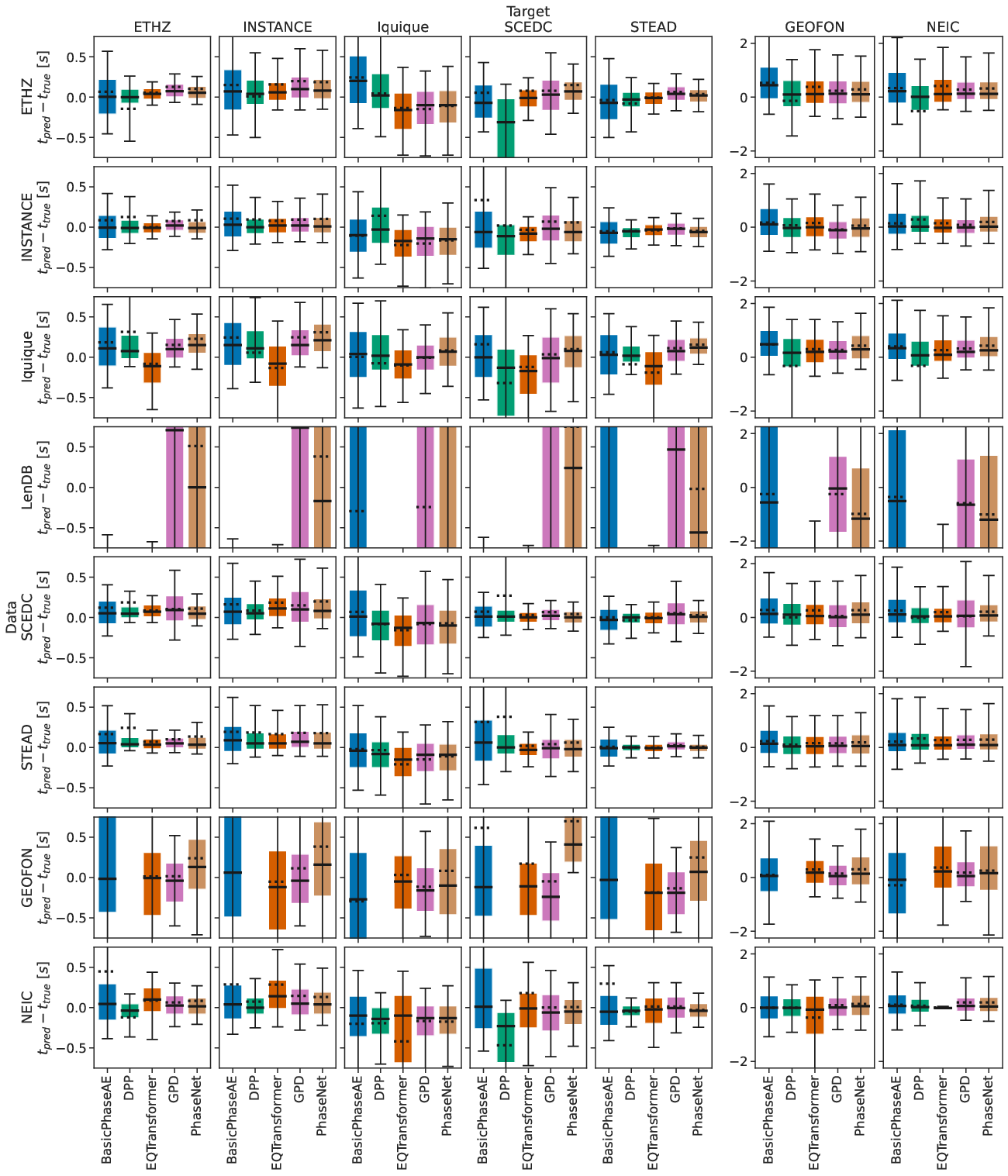}
    \caption{Distribution of S pick residuals from cross-domain experiments. Each panel shows one combination of training (row) and evaluation (column) dataset, each bar one model. The solid bars show the interquartile range, the whiskers range from the 10th to the 90th percentile. The solid lines indicate medians, the dashed lines indicate means.}
    \label{fig:s_diff_cross}
\end{figure*}

\begin{table*}
 \centering
 \caption{Phase identification results (BasicPhaseAE)}
\input{tables/cross/basicphaseae_phase_test.tex}
\label{tab:phase_test}
\end{table*}

\begin{table*}
 \centering
 \caption{Phase identification results (DPP)}
\input{tables/cross/dppdetect_phase_test.tex}
\label{tab:phase_test}
\end{table*}

\begin{table*}
 \centering
 \caption{Phase identification results (EQTransformer)}
\input{tables/cross/eqtransformer_phase_test.tex}
\label{tab:phase_test}
\end{table*}

\begin{table*}
 \centering
 \caption{Phase identification results (GPD-Org)}
\input{tables/cross/gpd_phase_test.tex}
\label{tab:phase_test}
\end{table*}

\begin{table*}
 \centering
 \caption{Phase identification results (GPD)}
\input{tables/cross/gpdpick_phase_test.tex}
\label{tab:phase_test}
\end{table*}

\begin{table*}
 \centering
 \caption{Phase identification results (PhaseNet)}
\input{tables/cross/phasenet_phase_test.tex}
\label{tab:phase_test}
\end{table*}

\begin{figure*}
    \centering
    \includegraphics[width=\textwidth]{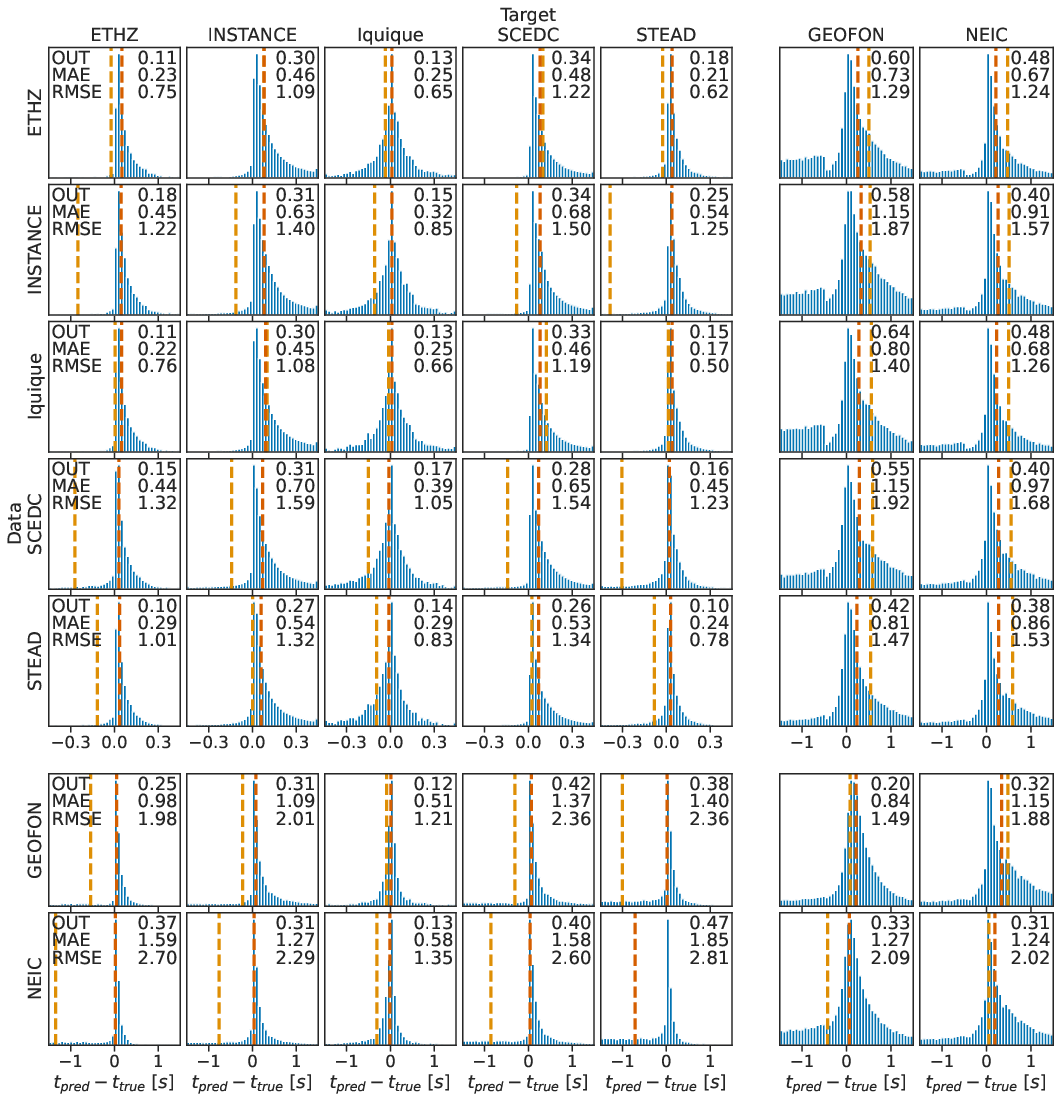}
    \caption{P residuals (Baer-Kradolfer). For detailed description see Figure 3 in the main text.}
    \label{fig:test_P_diff}
\end{figure*}

\begin{figure*}
    \centering
    \includegraphics[width=\textwidth]{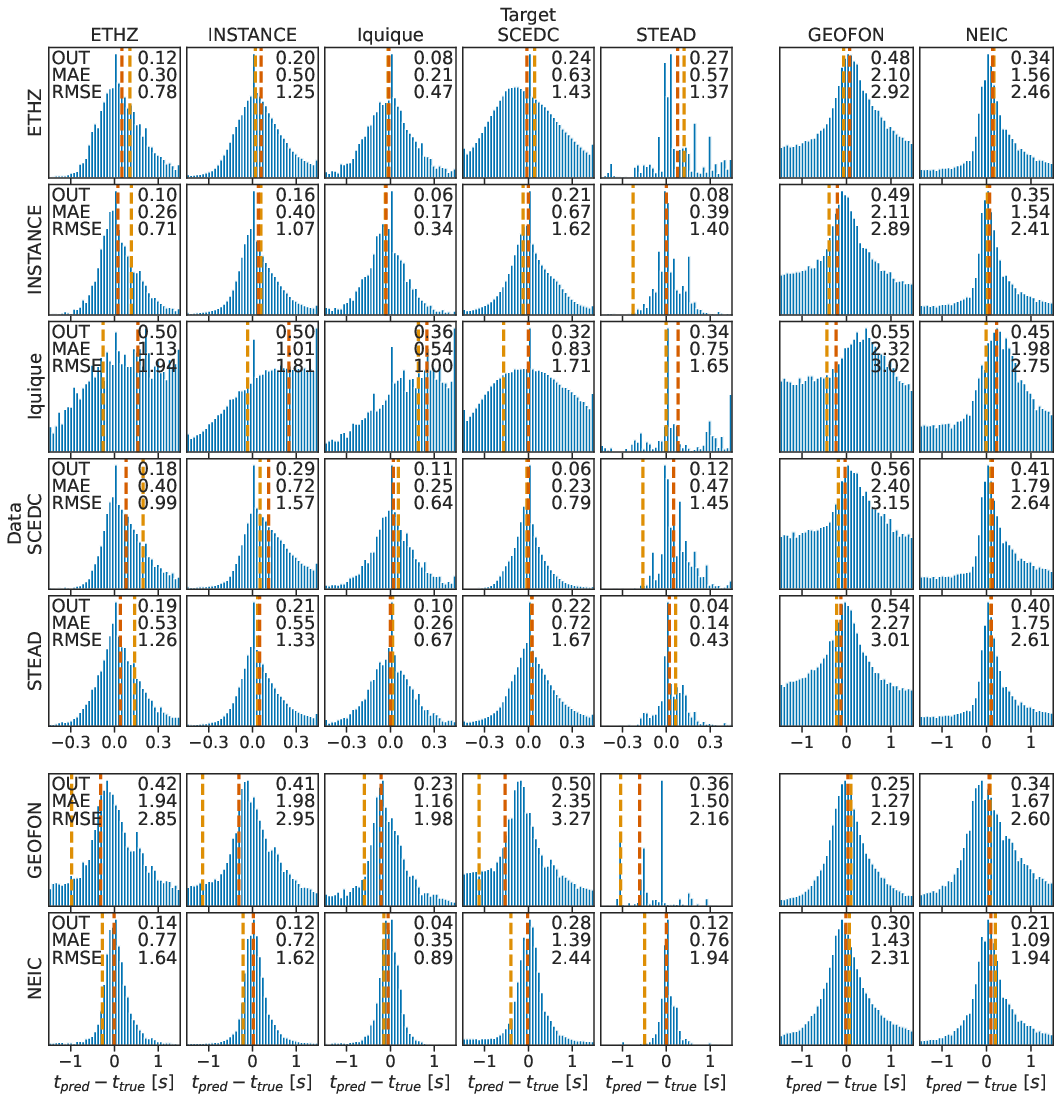}
    \caption{P residuals (BasicPhaseAE). For detailed description see Figure 3 in the main text.}
    \label{fig:test_P_diff}
\end{figure*}

\begin{figure*}
    \centering
    \includegraphics[width=\textwidth]{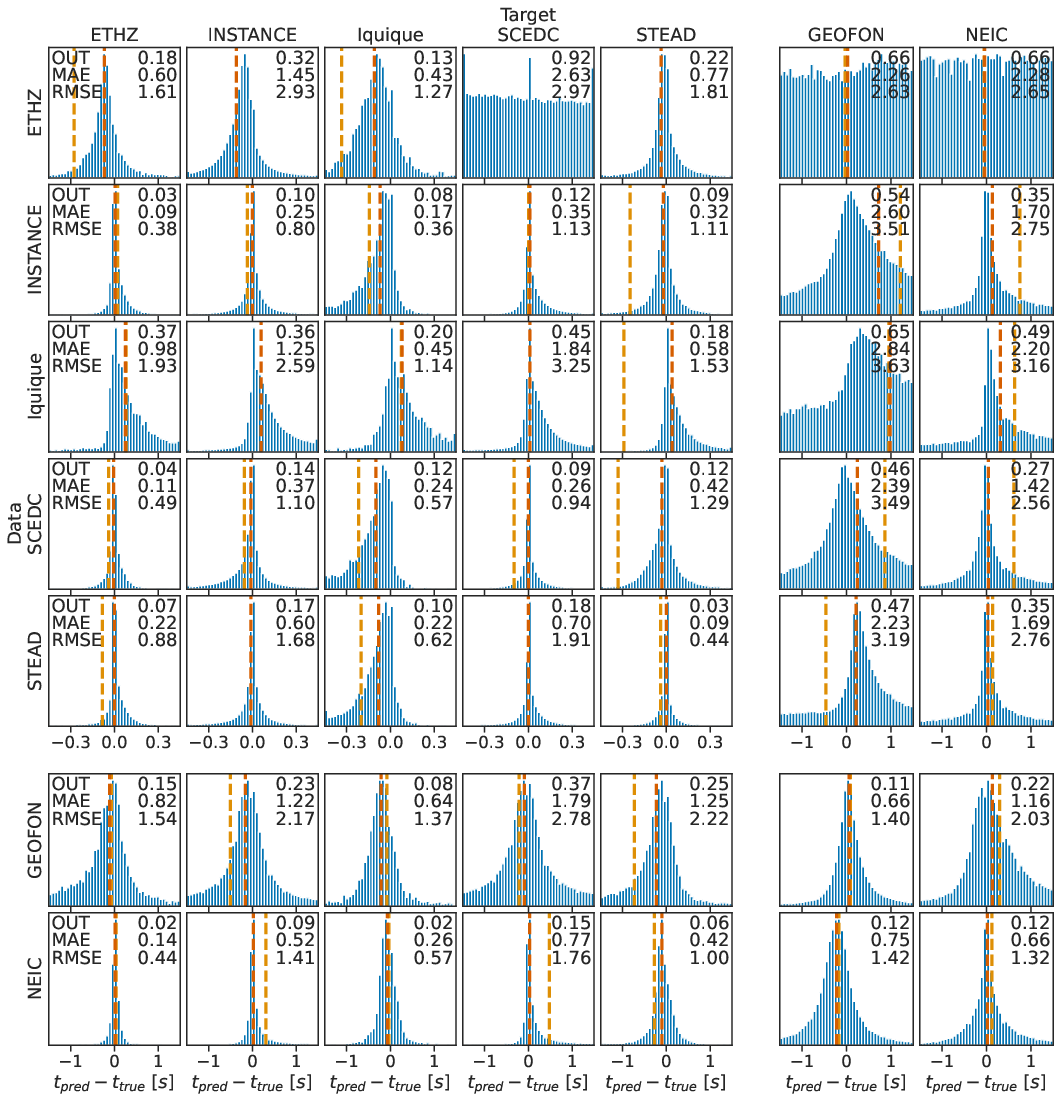}
    \caption{P residuals (DPP). For detailed description see Figure 3 in the main text.}
    \label{fig:test_P_diff}
\end{figure*}

\begin{figure*}
    \centering
    \includegraphics[width=\textwidth]{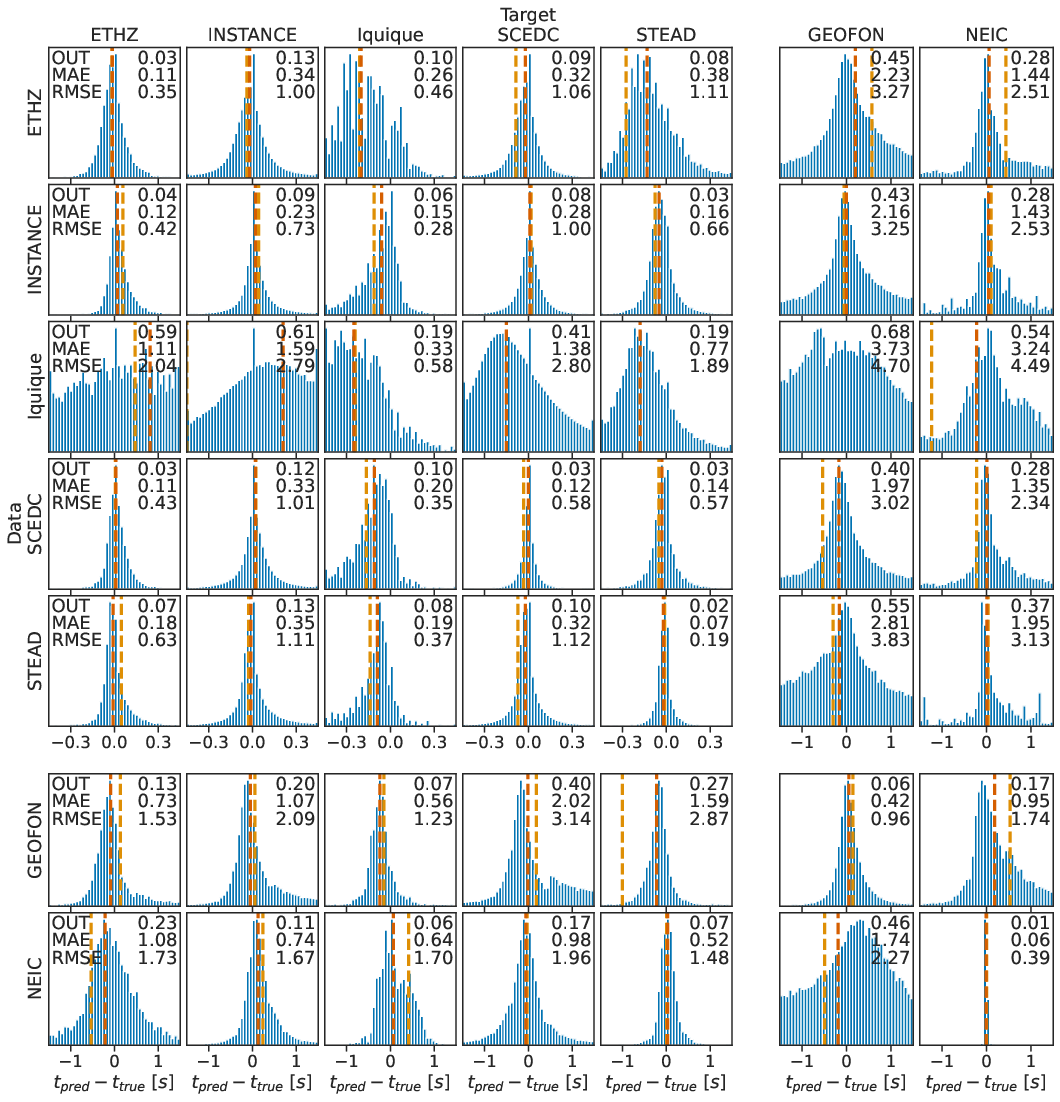}
    \caption{P residuals (EQTransformer). For detailed description see Figure 3 in the main text.}
    \label{fig:test_P_diff}
\end{figure*}

\begin{figure*}
    \centering
    \includegraphics[width=\textwidth]{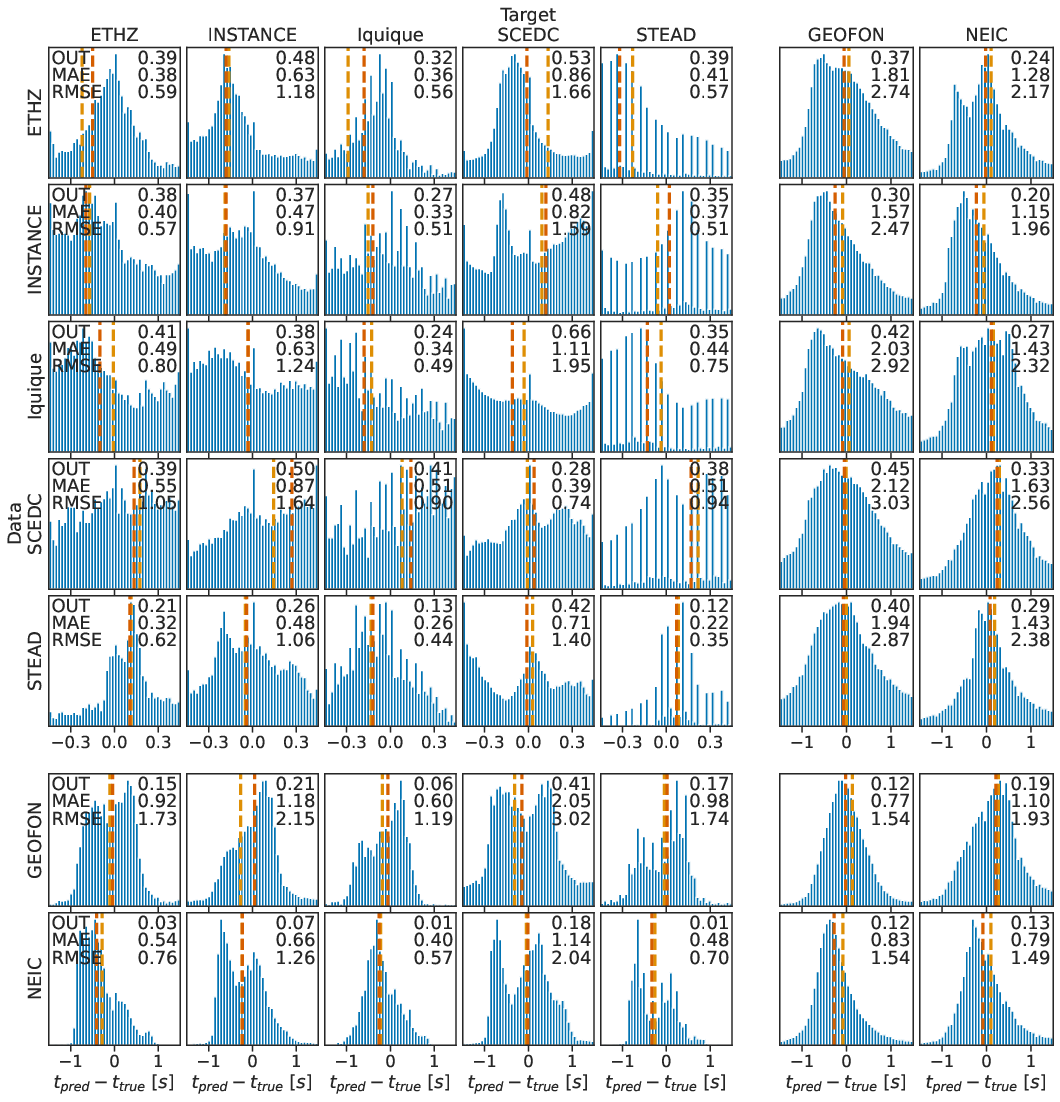}
    \caption{P residuals (GPD-Org). For detailed description see Figure 3 in the main text.}
    \label{fig:test_P_diff}
\end{figure*}

\begin{figure*}
    \centering
    \includegraphics[width=\textwidth]{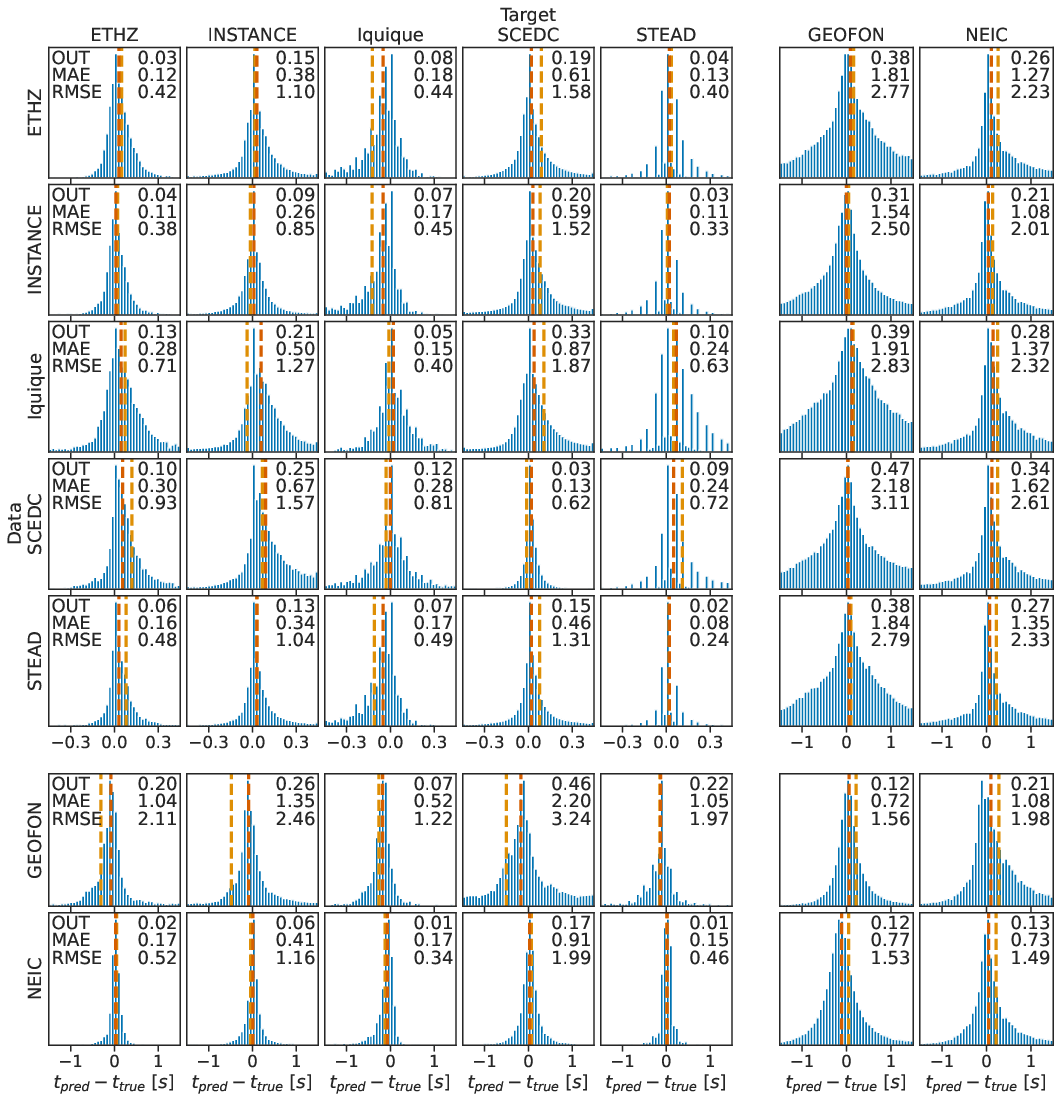}
    \caption{P residuals (GPD). For detailed description see Figure 3 in the main text.}
    \label{fig:test_P_diff}
\end{figure*}

\begin{figure*}
    \centering
    \includegraphics[width=\textwidth]{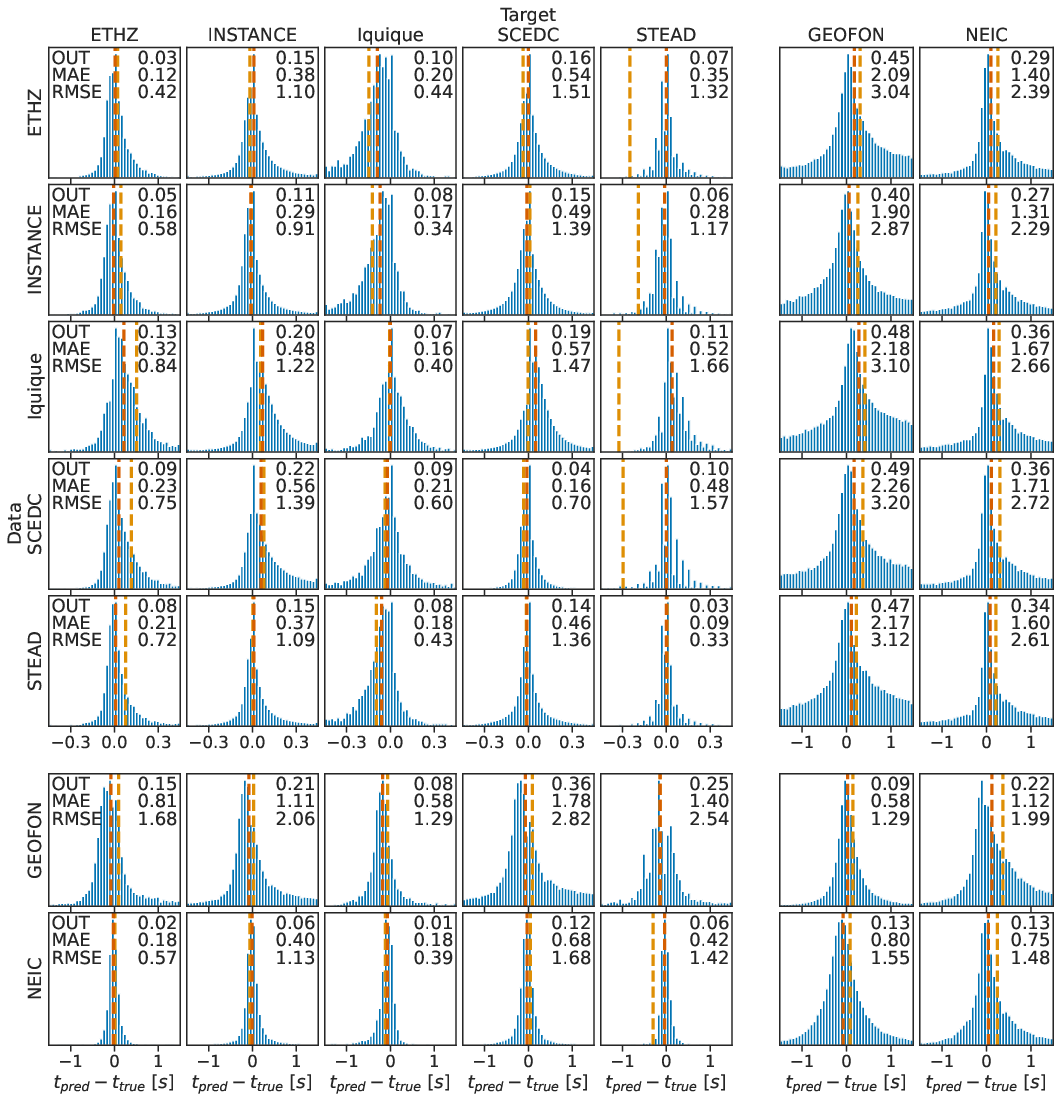}
    \caption{P residuals (PhaseNet). For detailed description see Figure 3 in the main text.}
    \label{fig:test_P_diff}
\end{figure*}

\begin{figure*}
    \centering
    \includegraphics[width=\textwidth]{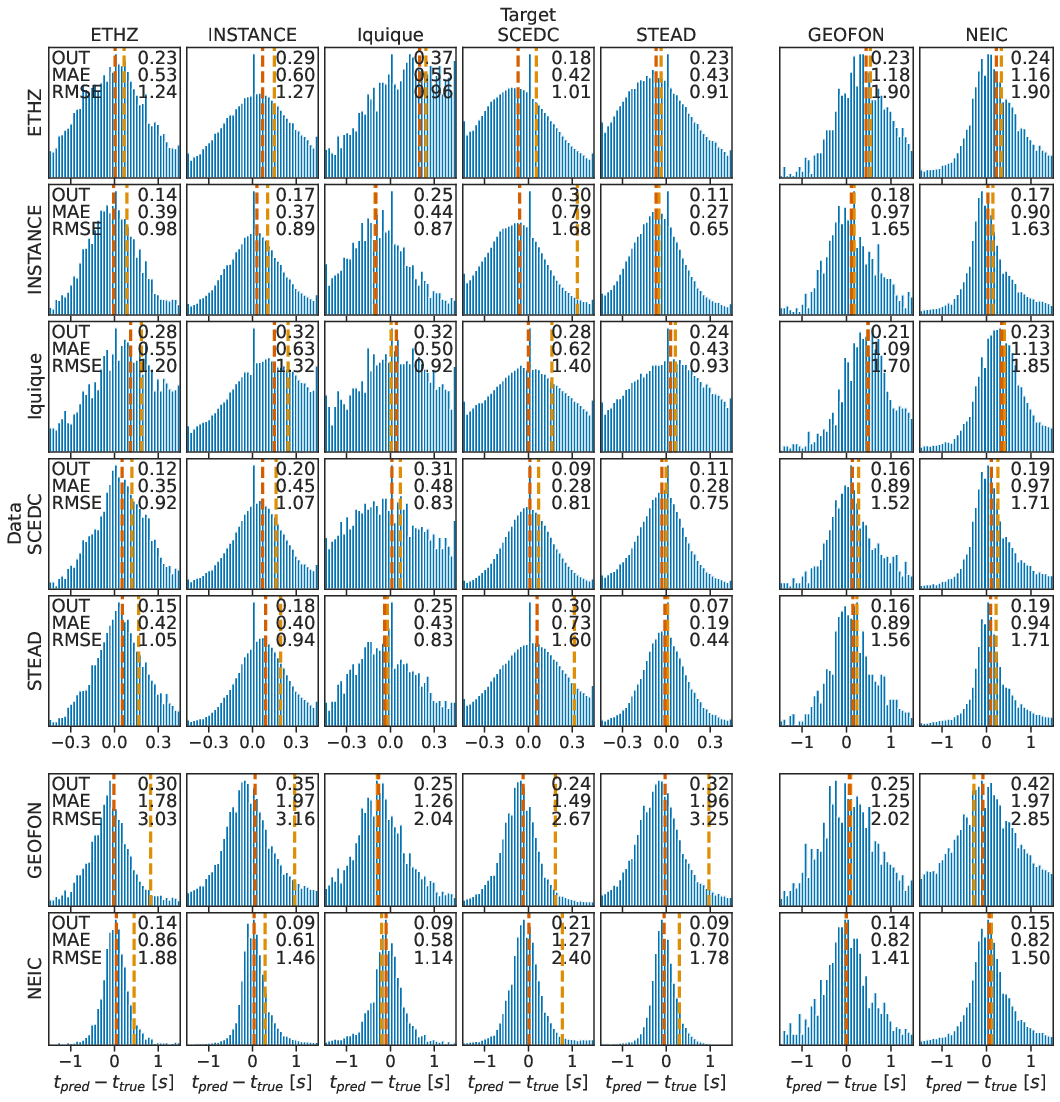}
    \caption{S residuals (BasicPhaseAE). For detailed description see Figure 4 in the main text.}
    \label{fig:test_P_diff}
\end{figure*}

\begin{figure*}
    \centering
    \includegraphics[width=\textwidth]{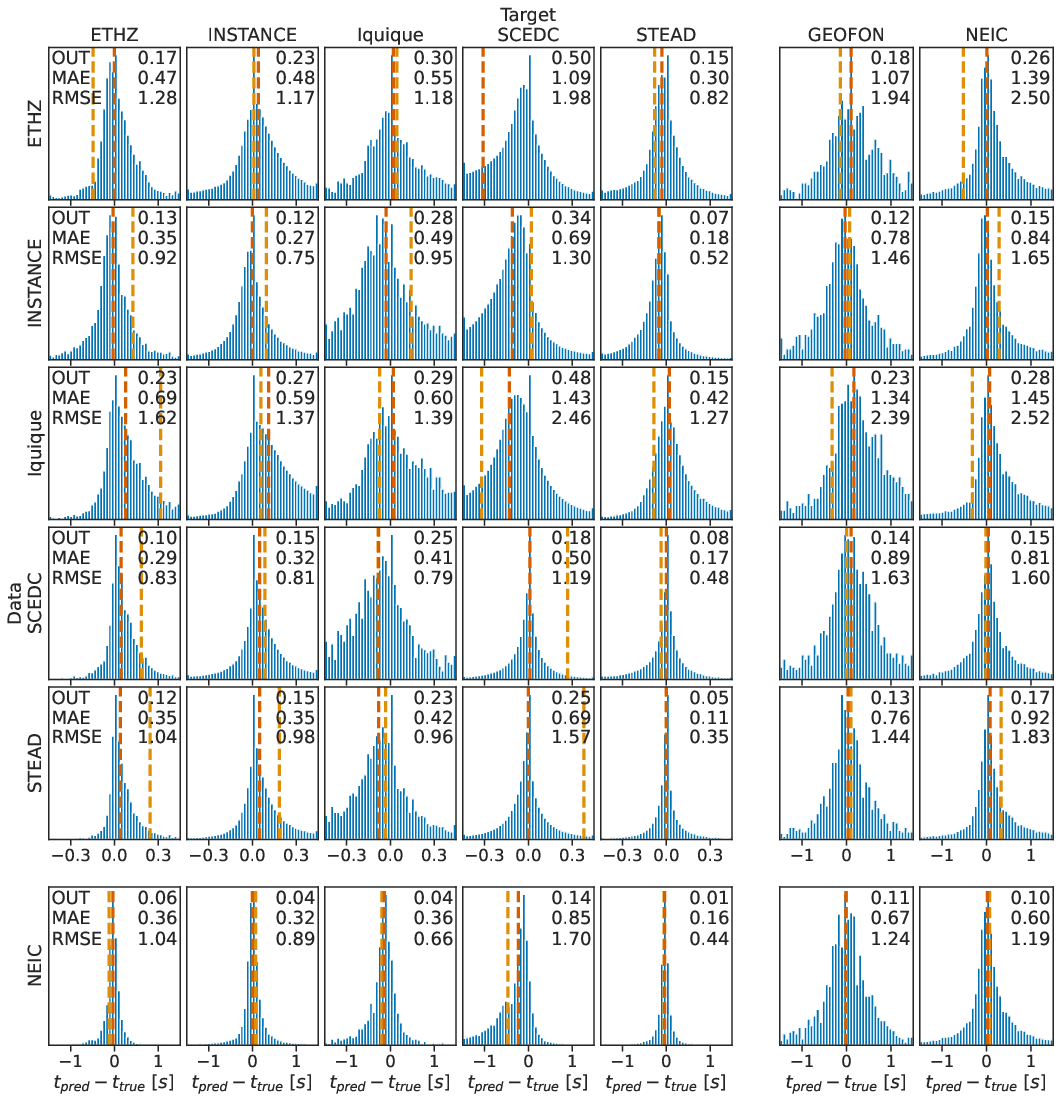}
    \caption{S residuals (DPP). For detailed description see Figure 4 in the main text.}
    \label{fig:test_P_diff}
\end{figure*}

\begin{figure*}
    \centering
    \includegraphics[width=\textwidth]{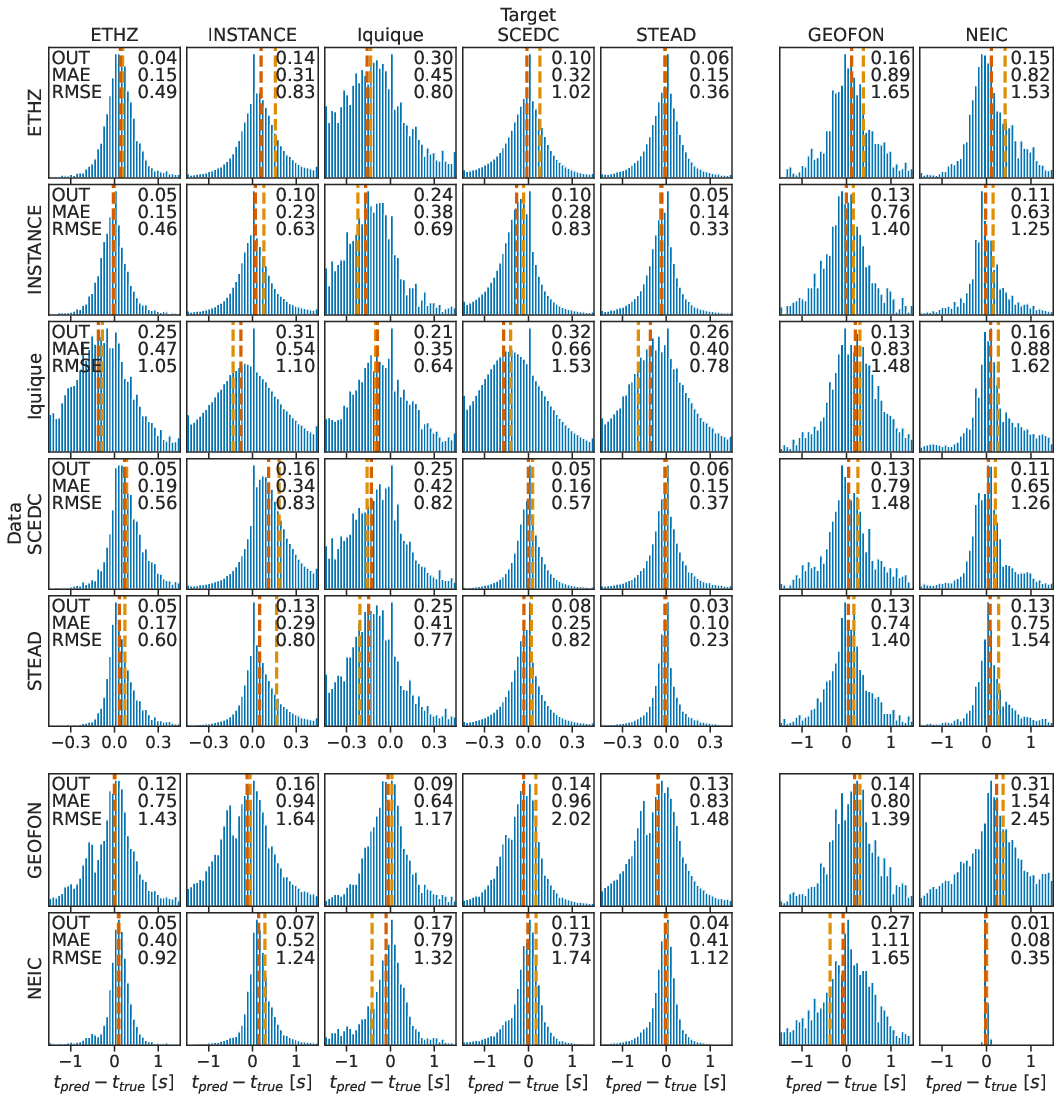}
    \caption{S residuals (EQTransformer). For detailed description see Figure 4 in the main text.}
    \label{fig:test_P_diff}
\end{figure*}

\begin{figure*}
    \centering
    \includegraphics[width=\textwidth]{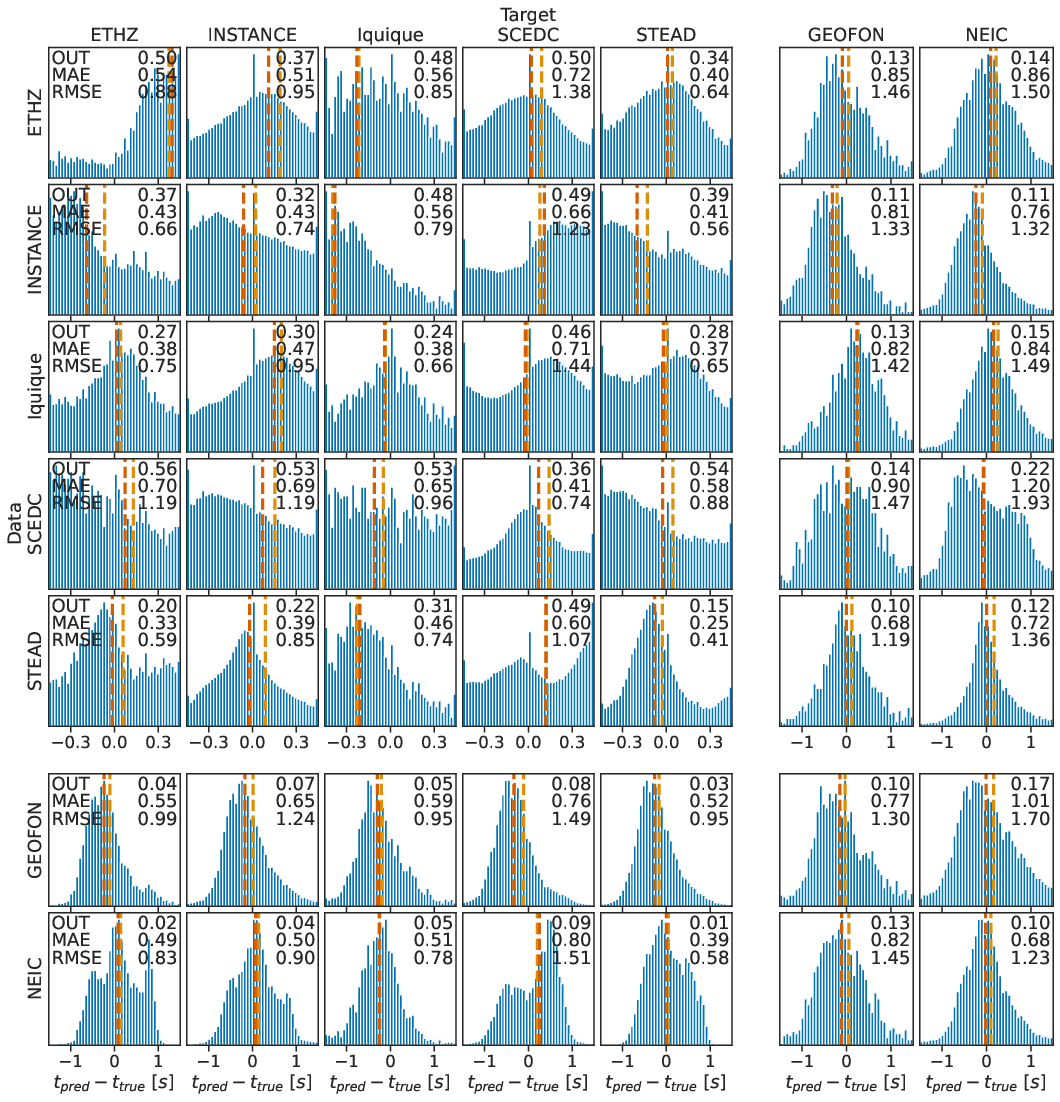}
    \caption{S residuals (GPD-Org). For detailed description see Figure 4 in the main text.}
    \label{fig:test_P_diff}
\end{figure*}

\begin{figure*}
    \centering
    \includegraphics[width=\textwidth]{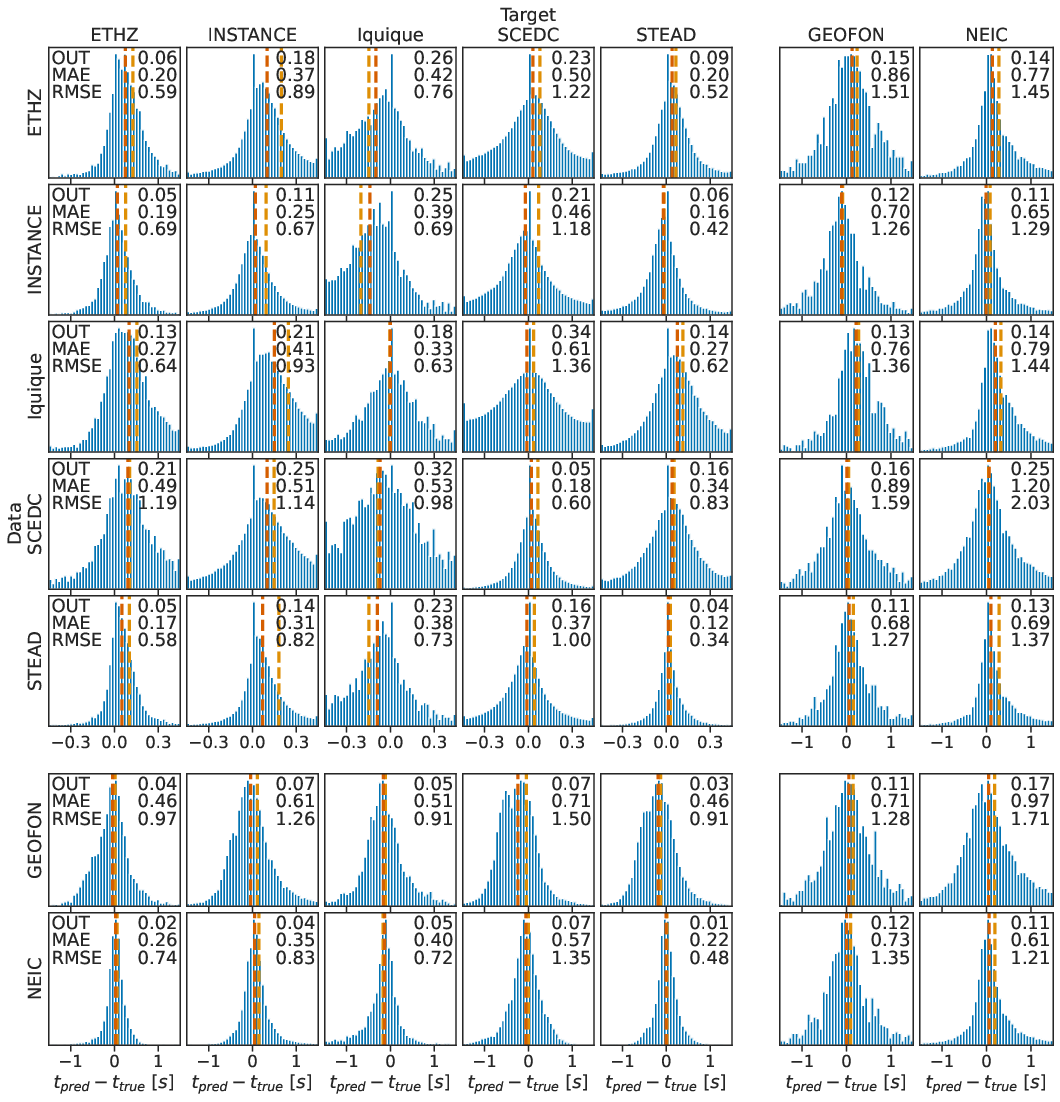}
    \caption{S residuals (GPD-Pick). For detailed description see Figure 4 in the main text.}
    \label{fig:test_P_diff}
\end{figure*}

\begin{figure*}
    \centering
    \includegraphics[width=\textwidth]{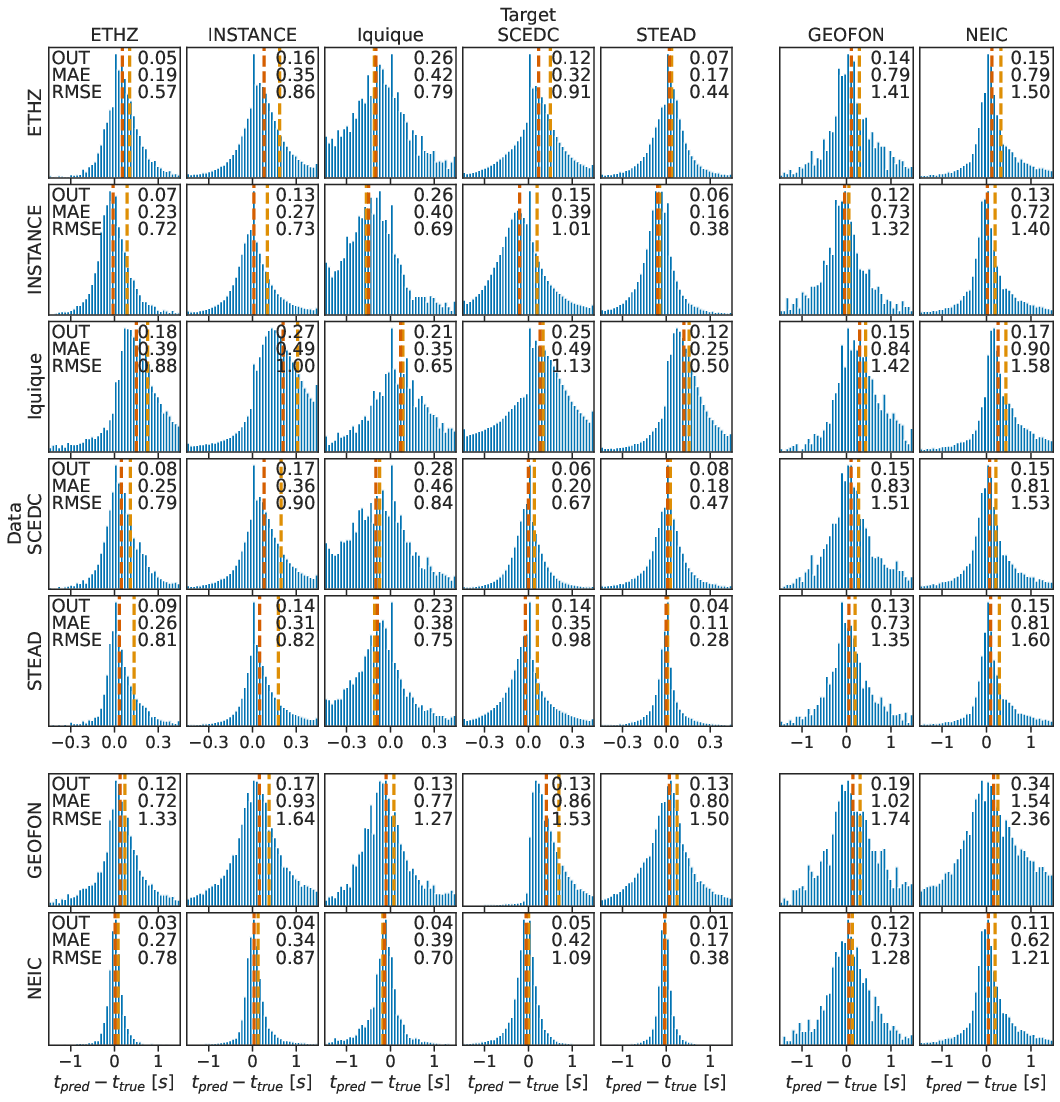}
    \caption{S residuals (PhaseNet). For detailed description see Figure 4 in the main text.}
    \label{fig:test_P_diff}
\end{figure*}

\begin{table*}
 \centering
 \small
 \caption{Detection results on GEOFON for resampled models}
\input{tables/resampled/detection_test_geofon.tex}
\label{tab:detection_test}
\end{table*}

\begin{table*}
 \centering
 \small
 \caption{Phase identification on GEOFON for resampled models}
\input{tables/resampled/phase_test_geofon.tex}
\label{tab:phase_test}
\end{table*}

\begin{table*}
 \centering
 \small
 \caption{Phase identification on NEIC for resampled models}
\input{tables/resampled/phase_test_neic.tex}
\label{tab:phase_test}
\end{table*}

\begin{figure*}
    \centering
    \includegraphics[width=\textwidth]{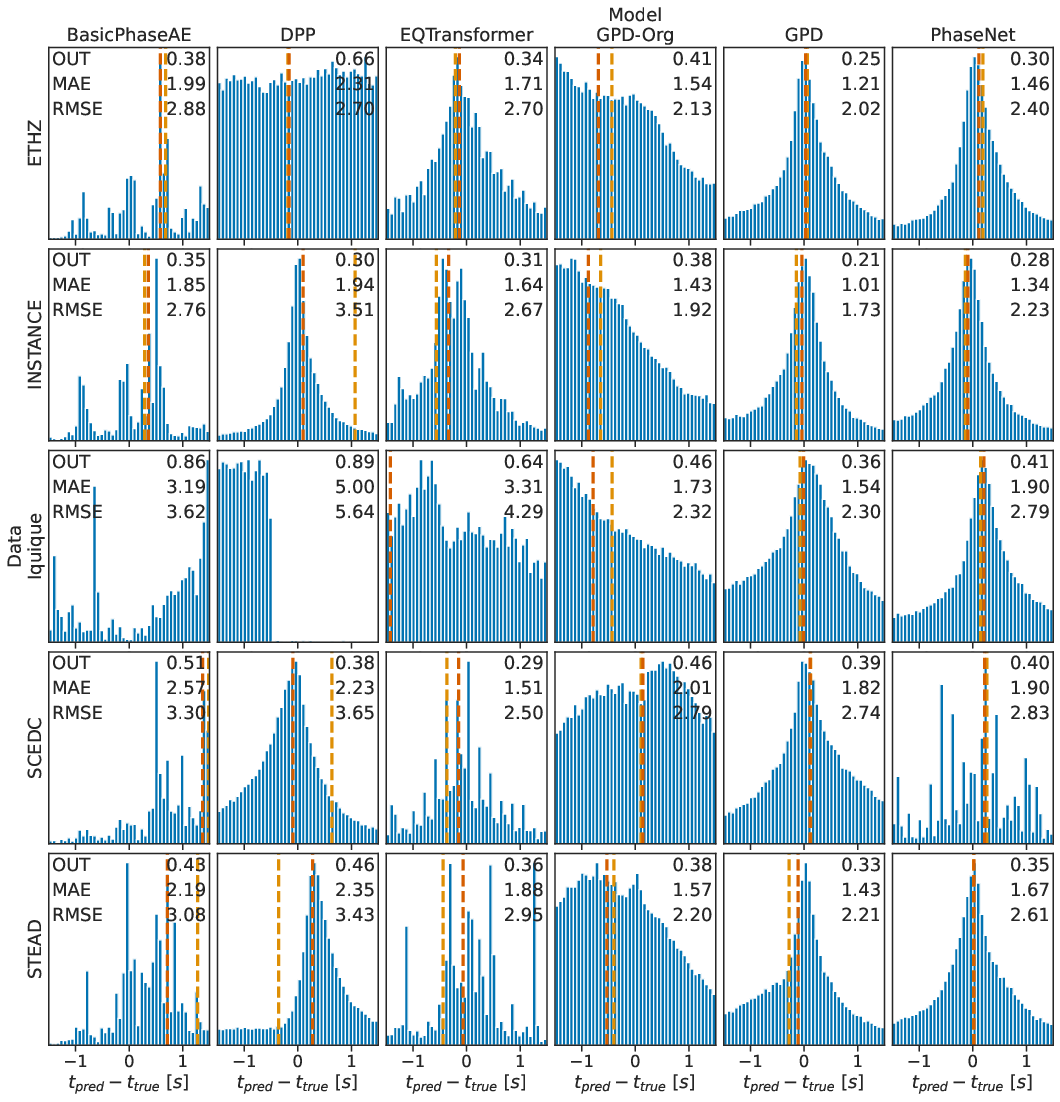}
    \caption{P residuals on GEOFON for resampled models. For detailed description see Figure 3 in the main text.}
    \label{fig:test_P_diff}
\end{figure*}

\begin{figure*}
    \centering
    \includegraphics[width=\textwidth]{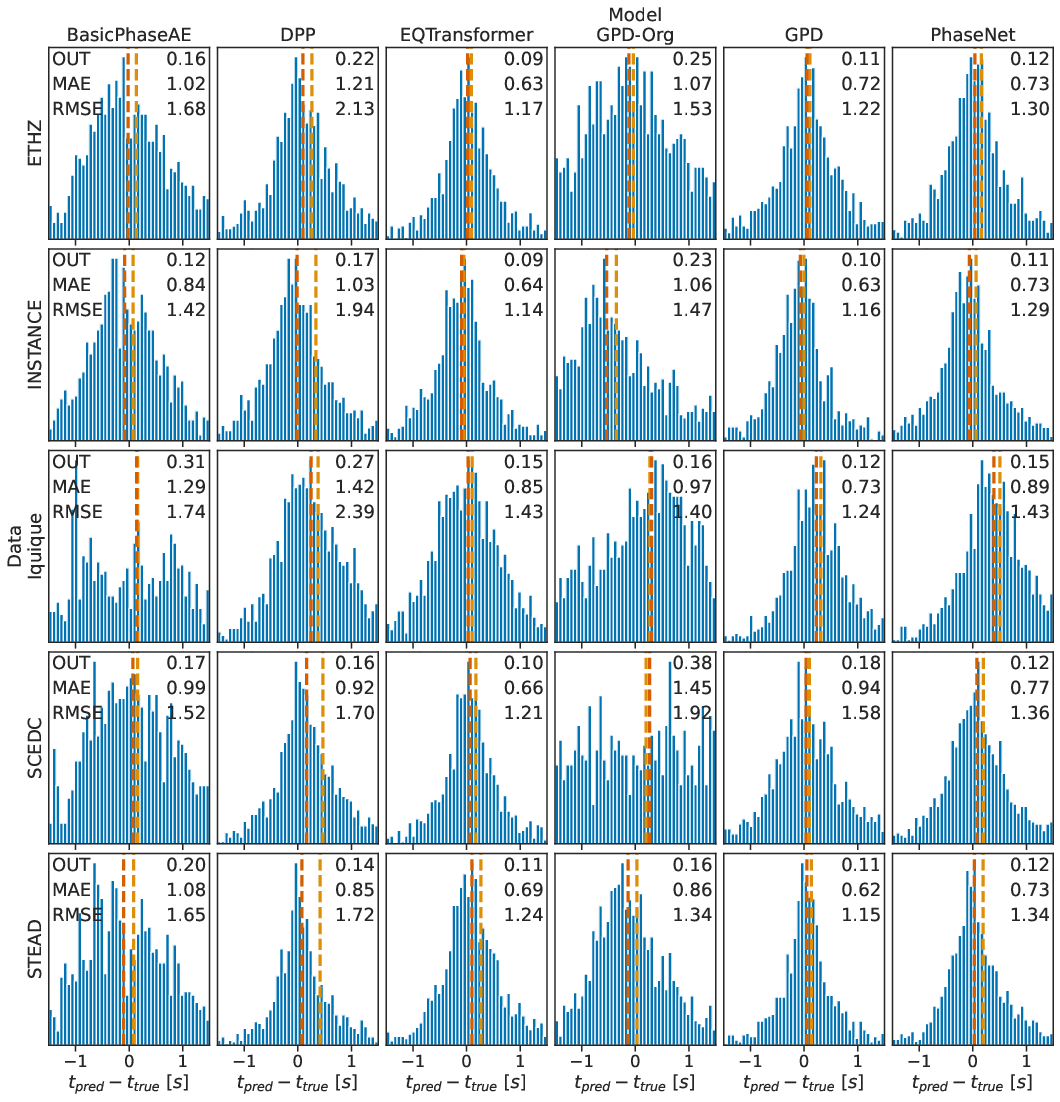}
    \caption{S residuals on GEOFON for resampled models. For detailed description see Figure 4 in the main text.}
    \label{fig:test_P_diff}
\end{figure*}

\begin{figure*}
    \centering
    \includegraphics[width=\textwidth]{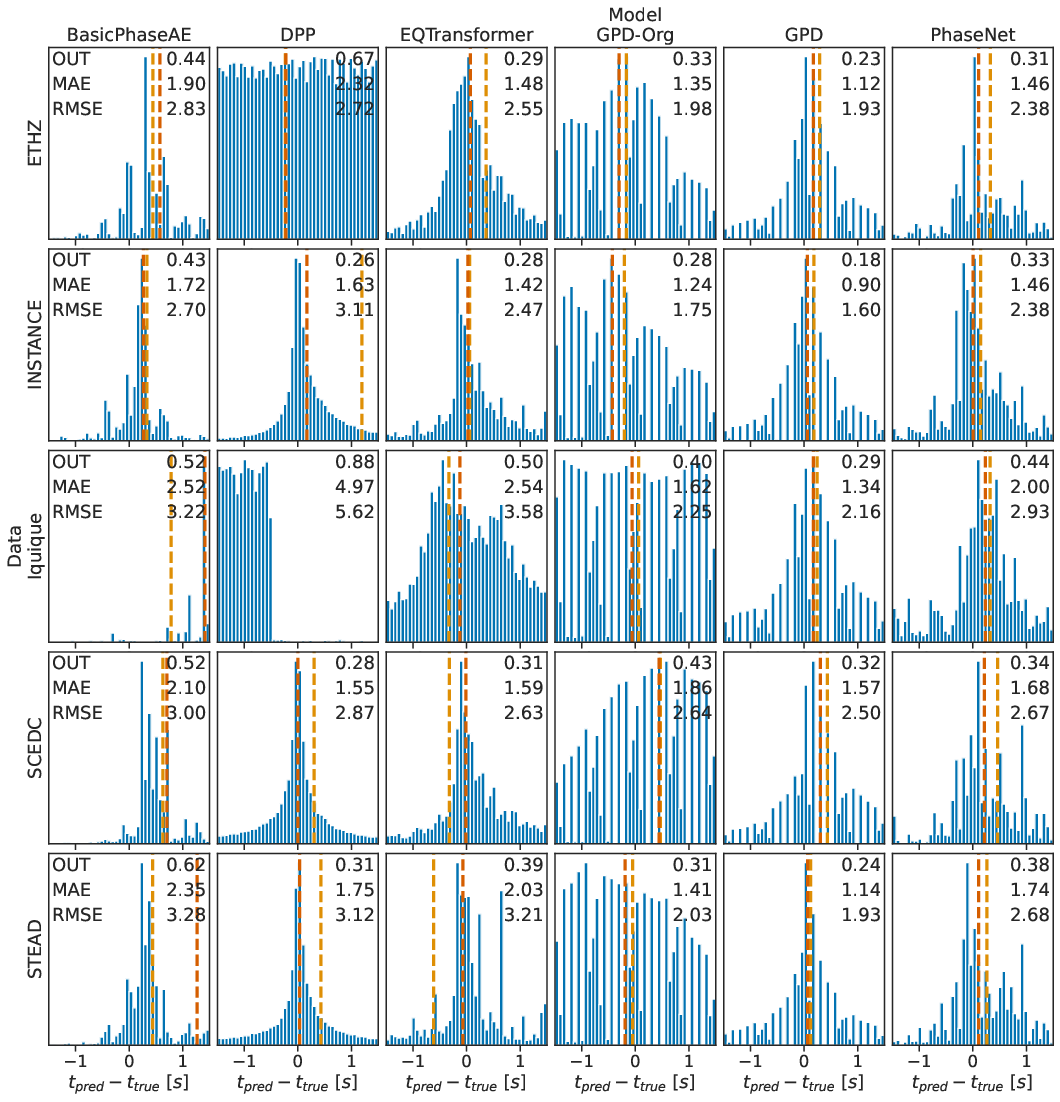}
    \caption{P residuals on NEIC for resampled models. For detailed description see Figure 3 in the main text.}
    \label{fig:test_P_diff}
\end{figure*}

\begin{figure*}
    \centering
    \includegraphics[width=\textwidth]{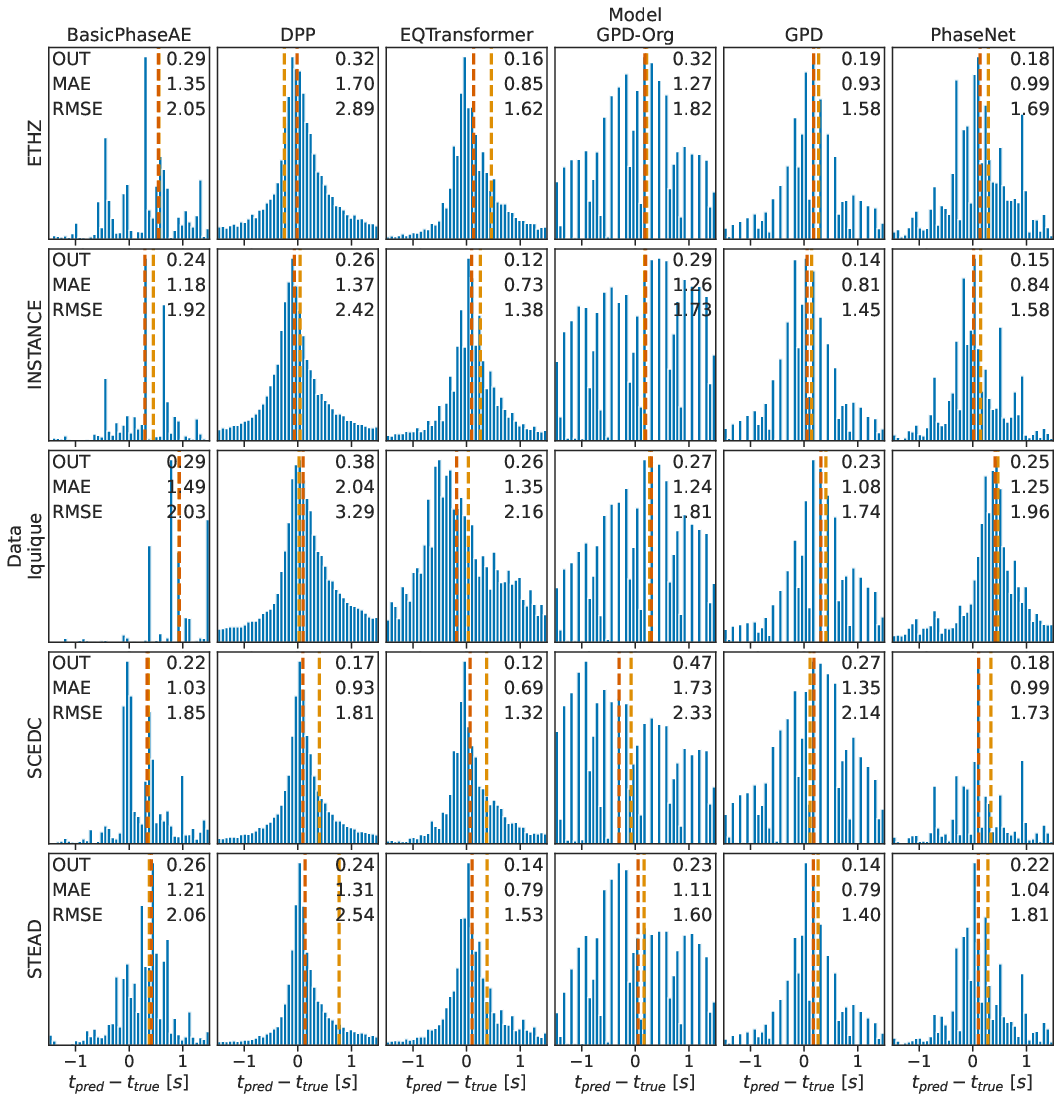}
    \caption{S residuals on NEIC for resampled models. For detailed description see Figure 4 in the main text.}
    \label{fig:test_P_diff}
\end{figure*}

\end{document}

%% file: tables/phase_test_mcc.tex
\setlength{\tabcolsep}{2pt}
\begin{tabular}{c|c|c|c|c|c|c}
\backslashbox{Data}{Model} & \multicolumn{1}{|c}{BasicPhaseAE} & \multicolumn{1}{|c}{DPP} & \multicolumn{1}{|c}{EQTransformer} & \multicolumn{1}{|c}{GPD} & \multicolumn{1}{|c}{PhaseNet} & \multicolumn{1}{|c}{$\diameter$} \\
 & MCC & MCC & MCC & MCC & MCC & MCC\\
\hline
ETHZ & 0.77 & 0.89 & 0.97 & 0.92 & 0.91 & 0.89 \\
INSTANCE & 0.87 & 0.89 & 0.97 & 0.95 & 0.94 & 0.92 \\
Iquique & 0.81 & 0.91 & 0.99 & 0.98 & 0.96 & 0.93 \\
SCEDC & 0.84 & 0.82 & 0.96 & 0.93 & 0.91 & 0.89 \\
STEAD & 0.92 & 0.57 & 1.00 & 0.99 & 0.99 & 0.89 \\
GEOFON & 0.06 & 0.46 & 0.82 & 0.67 & 0.51 & 0.50 \\
NEIC & 0.70 & 0.76 & 0.96 & 0.84 & 0.81 & 0.81 \\
\hline
$\diameter$ & 0.71 & 0.76 & 0.95 & 0.90 & 0.86 \\
\end{tabular}

%% file: tables/gpd/detection_test_gpd.tex
\setlength{\tabcolsep}{2pt}
\begin{tabular}{c|c|c|c|c|c|c|c|c}
\backslashbox{Data}{Model} & \multicolumn{1}{|c}{BasicPhaseAE} & \multicolumn{1}{|c}{CRED} & \multicolumn{1}{|c}{DPP} & \multicolumn{1}{|c}{EQTransformer} & \multicolumn{1}{|c}{GPD-Org} & \multicolumn{1}{|c}{GPD} & \multicolumn{1}{|c}{PhaseNet} & \multicolumn{1}{|c}{$\diameter$} \\
 & AUC & AUC & AUC & AUC & AUC & AUC & AUC & AUC\\
\hline
ETHZ & 0.94 & 0.96 & 0.98 & 0.96 & 0.98 & 0.99 & 0.98 & 0.97 \\
INSTANCE & 0.90 & 0.95 & 0.96 & 0.96 & 0.96 & 0.97 & 0.96 & 0.95 \\
LenDB & 0.30 & 0.98 & 0.94 & 0.98 & 0.94 & 0.95 & 0.95 & 0.86 \\
SCEDC & 0.89 & 0.86 & 0.92 & 0.90 & 0.78 & 0.90 & 0.92 & 0.88 \\
STEAD & 0.99 & 1.00 & 1.00 & 1.00 & 1.00 & 1.00 & 1.00 & 1.00 \\
GEOFON & 0.59 & 0.96 & 0.85 & 0.99 & 0.91 & 0.90 & 0.92 & 0.87 \\
\hline
$\diameter$ & 0.77 & 0.95 & 0.94 & 0.96 & 0.93 & 0.95 & 0.96 \\
\end{tabular}

%% file: tables/gpd/phase_test_mcc_gpd.tex
\setlength{\tabcolsep}{2pt}
\begin{tabular}{c|c|c|c|c|c|c|c}
\backslashbox{Data}{Model} & \multicolumn{1}{|c}{BasicPhaseAE} & \multicolumn{1}{|c}{DPP} & \multicolumn{1}{|c}{EQTransformer} & \multicolumn{1}{|c}{GPD-Org} & \multicolumn{1}{|c}{GPD} & \multicolumn{1}{|c}{PhaseNet} & \multicolumn{1}{|c}{$\diameter$} \\
 & MCC & MCC & MCC & MCC & MCC & MCC & MCC\\
\hline
ETHZ & 0.77 & 0.89 & 0.97 & 0.88 & 0.92 & 0.91 & 0.89 \\
INSTANCE & 0.87 & 0.89 & 0.97 & 0.90 & 0.95 & 0.94 & 0.92 \\
Iquique & 0.81 & 0.91 & 0.99 & 0.97 & 0.98 & 0.96 & 0.94 \\
SCEDC & 0.84 & 0.82 & 0.96 & 0.88 & 0.93 & 0.91 & 0.89 \\
STEAD & 0.92 & 0.57 & 1.00 & 0.95 & 0.99 & 0.99 & 0.90 \\
GEOFON & 0.06 & 0.46 & 0.82 & 0.66 & 0.67 & 0.51 & 0.53 \\
NEIC & 0.70 & 0.76 & 0.96 & 0.84 & 0.84 & 0.81 & 0.82 \\
\hline
$\diameter$ & 0.71 & 0.76 & 0.95 & 0.87 & 0.90 & 0.86 \\
\end{tabular}

%% file: tables/gpd/precision_p_test_gpd.tex
\setlength{\tabcolsep}{2pt}
\begin{tabular}{c|ccc|ccc|ccc|ccc|ccc|ccc|cc}
\backslashbox{Data}{Model} & \multicolumn{3}{|c}{BasicPhaseAE} & \multicolumn{3}{|c}{DPP} & \multicolumn{3}{|c}{EQTransformer} & \multicolumn{3}{|c}{GPD-Org} & \multicolumn{3}{|c}{GPD} & \multicolumn{3}{|c}{PhaseNet} & \multicolumn{2}{|c}{$\diameter$} \\
 & $\mu$ & $\sigma$ & MAE & $\mu$ & $\sigma$ & MAE & $\mu$ & $\sigma$ & MAE & $\mu$ & $\sigma$ & MAE & $\mu$ & $\sigma$ & MAE & $\mu$ & $\sigma$ & MAE & $\sigma$ & MAE\\
\hline
ETHZ & 0.11 & 0.78 & 0.30 & -0.28 & 1.61 & 0.60 & -0.02 & 0.35 & 0.11 & -0.22 & 0.59 & 0.38 & 0.05 & 0.42 & 0.12 & 0.02 & 0.42 & 0.12 & 0.70 & 0.27 \\
INSTANCE & 0.06 & 1.07 & 0.40 & -0.04 & 0.80 & 0.25 & 0.04 & 0.73 & 0.23 & -0.19 & 0.91 & 0.47 & -0.02 & 0.85 & 0.26 & -0.01 & 0.91 & 0.29 & 0.88 & 0.32 \\
Iquique & 0.19 & 1.00 & 0.54 & 0.07 & 1.14 & 0.45 & -0.24 & 0.58 & 0.33 & -0.13 & 0.49 & 0.34 & -0.01 & 0.40 & 0.15 & -0.00 & 0.40 & 0.16 & 0.67 & 0.33 \\
SCEDC & -0.01 & 0.79 & 0.23 & -0.10 & 0.94 & 0.26 & -0.03 & 0.58 & 0.12 & -0.01 & 0.74 & 0.39 & -0.01 & 0.62 & 0.13 & -0.03 & 0.70 & 0.16 & 0.73 & 0.22 \\
STEAD & 0.06 & 0.43 & 0.14 & -0.04 & 0.44 & 0.09 & -0.01 & 0.19 & 0.07 & 0.08 & 0.35 & 0.22 & 0.02 & 0.24 & 0.08 & 0.01 & 0.33 & 0.09 & 0.33 & 0.11 \\
GEOFON & 0.10 & 2.19 & 1.27 & 0.09 & 1.40 & 0.66 & 0.14 & 0.96 & 0.42 & 0.13 & 1.54 & 0.77 & 0.21 & 1.56 & 0.72 & 0.14 & 1.29 & 0.58 & 1.49 & 0.74 \\
NEIC & 0.20 & 1.94 & 1.09 & 0.12 & 1.32 & 0.66 & 0.00 & 0.39 & 0.06 & 0.10 & 1.49 & 0.79 & 0.22 & 1.49 & 0.73 & 0.24 & 1.48 & 0.75 & 1.35 & 0.68 \\
\hline
$\diameter$ & & 1.17 & 0.57 & & 1.09 & 0.42 & & 0.54 & 0.19 & & 0.87 & 0.48 & & 0.80 & 0.31 & & 0.79 & 0.31 \\
\end{tabular}

%% file: tables/gpd/precision_s_test_gpd.tex
\setlength{\tabcolsep}{2pt}
\begin{tabular}{c|ccc|ccc|ccc|ccc|ccc|ccc|cc}
\backslashbox{Data}{Model} & \multicolumn{3}{|c}{BasicPhaseAE} & \multicolumn{3}{|c}{DPP} & \multicolumn{3}{|c}{EQTransformer} & \multicolumn{3}{|c}{GPD-Org} & \multicolumn{3}{|c}{GPD} & \multicolumn{3}{|c}{PhaseNet} & \multicolumn{2}{|c}{$\diameter$} \\
 & $\mu$ & $\sigma$ & MAE & $\mu$ & $\sigma$ & MAE & $\mu$ & $\sigma$ & MAE & $\mu$ & $\sigma$ & MAE & $\mu$ & $\sigma$ & MAE & $\mu$ & $\sigma$ & MAE & $\sigma$ & MAE\\
\hline
ETHZ & 0.07 & 1.24 & 0.53 & -0.15 & 1.28 & 0.47 & 0.06 & 0.49 & 0.15 & 0.37 & 0.88 & 0.54 & 0.13 & 0.59 & 0.20 & 0.10 & 0.57 & 0.19 & 0.84 & 0.35 \\
INSTANCE & 0.10 & 0.89 & 0.37 & 0.10 & 0.75 & 0.27 & 0.08 & 0.63 & 0.23 & 0.02 & 0.74 & 0.43 & 0.09 & 0.67 & 0.25 & 0.10 & 0.73 & 0.27 & 0.74 & 0.30 \\
Iquique & 0.00 & 0.92 & 0.50 & -0.07 & 1.39 & 0.60 & -0.10 & 0.64 & 0.35 & -0.03 & 0.66 & 0.38 & -0.00 & 0.63 & 0.33 & 0.08 & 0.65 & 0.35 & 0.82 & 0.42 \\
SCEDC & 0.07 & 0.81 & 0.28 & 0.27 & 1.19 & 0.50 & 0.03 & 0.57 & 0.16 & 0.14 & 0.74 & 0.41 & 0.06 & 0.60 & 0.18 & 0.04 & 0.67 & 0.20 & 0.76 & 0.29 \\
STEAD & 0.01 & 0.44 & 0.19 & 0.00 & 0.35 & 0.11 & -0.00 & 0.23 & 0.10 & -0.03 & 0.41 & 0.25 & 0.03 & 0.34 & 0.12 & 0.01 & 0.28 & 0.11 & 0.34 & 0.15 \\
GEOFON & 0.09 & 2.02 & 1.25 & nan & nan & nan & 0.30 & 1.39 & 0.80 & -0.03 & 1.30 & 0.77 & 0.15 & 1.28 & 0.71 & 0.30 & 1.74 & 1.02 & 1.54 & 0.91 \\
NEIC & 0.11 & 1.50 & 0.82 & 0.07 & 1.19 & 0.60 & 0.00 & 0.35 & 0.08 & 0.10 & 1.23 & 0.68 & 0.19 & 1.21 & 0.61 & 0.19 & 1.21 & 0.62 & 1.12 & 0.57 \\
\hline
$\diameter$ & & 1.12 & 0.57 & & 1.02 & 0.43 & & 0.61 & 0.27 & & 0.85 & 0.49 & & 0.76 & 0.34 & & 0.84 & 0.39 \\
\end{tabular}

%% file: tables/detection_thresholds.tex
\setlength{\tabcolsep}{2pt}
\begin{tabular}{c|c|c|c|c|c|c|c|c}
\backslashbox{Data}{Model} & \multicolumn{1}{|c}{BasicPhaseAE} & \multicolumn{1}{|c}{CRED} & \multicolumn{1}{|c}{DPP} & \multicolumn{1}{|c}{EQTransformer} & \multicolumn{1}{|c}{GPD-Org} & \multicolumn{1}{|c}{GPD} & \multicolumn{1}{|c}{PhaseNet} & \multicolumn{1}{|c}{$\diameter$} \\
 & Thr & Thr & Thr & Thr & Thr & Thr & Thr & Thr\\
\hline
ETHZ & 0.30 & 0.06 & 0.49 & 0.08 & 0.94 & 0.82 & 0.45 & 0.45 \\
INSTANCE & 0.04 & 0.01 & 0.29 & 0.01 & 0.66 & 0.48 & 0.11 & 0.23 \\
LenDB & 0.06 & 0.58 & 0.52 & 0.58 & 0.69 & 0.42 & 0.30 & 0.45 \\
SCEDC & 0.23 & 0.04 & 0.49 & 0.01 & 0.89 & 0.86 & 0.35 & 0.41 \\
STEAD & 0.38 & 0.27 & 0.54 & 0.72 & 0.95 & 0.76 & 0.49 & 0.59 \\
GEOFON & 0.03 & 0.67 & 0.61 & 0.88 & 0.82 & 0.58 & 0.18 & 0.54 \\
\hline
$\diameter$ & 0.17 & 0.27 & 0.49 & 0.38 & 0.83 & 0.65 & 0.31 \\
\end{tabular}

%% file: tables/phase_thresholds.tex
\setlength{\tabcolsep}{2pt}
\begin{tabular}{c|c|c|c|c|c|c|c}
\backslashbox{Data}{Model} & \multicolumn{1}{|c}{BasicPhaseAE} & \multicolumn{1}{|c}{DPP} & \multicolumn{1}{|c}{EQTransformer} & \multicolumn{1}{|c}{GPD-Org} & \multicolumn{1}{|c}{GPD} & \multicolumn{1}{|c}{PhaseNet} & \multicolumn{1}{|c}{$\diameter$} \\
 & Thr & Thr & Thr & Thr & Thr & Thr & Thr\\
\hline
ETHZ & 2.62 & 1.01 & 1.08 & 1.00 & 0.98 & 0.99 & 1.28 \\
INSTANCE & 1.44 & 0.37 & 0.70 & 1.01 & 0.89 & 0.86 & 0.88 \\
Iquique & 1.62 & 0.32 & 0.18 & 0.78 & 0.81 & 1.06 & 0.79 \\
SCEDC & 1.29 & 0.79 & 1.02 & 1.01 & 0.99 & 0.83 & 0.99 \\
STEAD & 1.77 & 0.22 & 1.01 & 1.00 & 0.98 & 1.06 & 1.01 \\
GEOFON & 46.38 & 1.14 & 3.65 & 1.62 & 1.95 & 1.94 & 9.44 \\
NEIC & 0.31 & 0.78 & 0.91 & 0.98 & 0.89 & 0.53 & 0.73 \\
\hline
$\diameter$ & 7.92 & 0.66 & 1.22 & 1.06 & 1.07 & 1.04 \\
\end{tabular}

%% file: tables/cross/basicphaseae_phase_test.tex
\setlength{\tabcolsep}{2pt}
\begin{tabular}{c|c|c|c|c|c|c|c|c}
\backslashbox{Data}{Target} & \multicolumn{1}{|c}{ETHZ} & \multicolumn{1}{|c}{INSTANCE} & \multicolumn{1}{|c}{Iquique} & \multicolumn{1}{|c}{SCEDC} & \multicolumn{1}{|c}{STEAD} & \multicolumn{1}{|c}{GEOFON} & \multicolumn{1}{|c}{NEIC} & \multicolumn{1}{|c}{$\diameter$} \\
 & MCC & MCC & MCC & MCC & MCC & MCC & MCC & MCC\\
\hline
ETHZ & 0.77 & 0.78 & 0.88 & 0.49 & 0.84 & 0.22 & 0.48 & 0.64 \\
INSTANCE & 0.79 & 0.87 & 0.89 & 0.36 & 0.90 & 0.35 & 0.61 & 0.68 \\
Iquique & 0.59 & 0.63 & 0.81 & 0.56 & 0.69 & 0.20 & 0.44 & 0.56 \\
SCEDC & 0.58 & 0.56 & 0.70 & 0.84 & 0.67 & 0.17 & 0.33 & 0.55 \\
STEAD & 0.71 & 0.82 & 0.85 & 0.39 & 0.92 & 0.48 & 0.64 & 0.69 \\
GEOFON & 0.09 & 0.05 & 0.03 & 0.18 & -0.00 & 0.06 & 0.08 & 0.07 \\
NEIC & 0.62 & 0.73 & 0.84 & 0.19 & 0.72 & 0.59 & 0.70 & 0.63 \\
\hline
$\diameter$ & 0.59 & 0.63 & 0.71 & 0.43 & 0.68 & 0.30 & 0.47 \\
\end{tabular}

%% file: tables/cross/dppdetect_phase_test.tex
\setlength{\tabcolsep}{2pt}
\begin{tabular}{c|c|c|c|c|c|c|c|c}
\backslashbox{Data}{Target} & \multicolumn{1}{|c}{ETHZ} & \multicolumn{1}{|c}{INSTANCE} & \multicolumn{1}{|c}{Iquique} & \multicolumn{1}{|c}{SCEDC} & \multicolumn{1}{|c}{STEAD} & \multicolumn{1}{|c}{GEOFON} & \multicolumn{1}{|c}{NEIC} & \multicolumn{1}{|c}{$\diameter$} \\
 & MCC & MCC & MCC & MCC & MCC & MCC & MCC & MCC\\
\hline
ETHZ & 0.89 & 0.84 & 0.91 & 0.52 & 0.59 & 0.39 & 0.65 & 0.68 \\
INSTANCE & 0.88 & 0.89 & 0.92 & 0.52 & 0.62 & 0.51 & 0.65 & 0.71 \\
Iquique & 0.74 & 0.77 & 0.91 & 0.42 & 0.62 & 0.40 & 0.63 & 0.64 \\
SCEDC & 0.62 & 0.61 & 0.74 & 0.82 & 0.45 & 0.23 & 0.38 & 0.55 \\
STEAD & 0.81 & 0.83 & 0.87 & 0.49 & 0.57 & 0.51 & 0.63 & 0.67 \\
GEOFON & 0.41 & 0.46 & 0.59 & 0.21 & 0.21 & 0.46 & 0.45 & 0.40 \\
NEIC & 0.74 & 0.78 & 0.88 & 0.27 & 0.59 & 0.65 & 0.76 & 0.67 \\
\hline
$\diameter$ & 0.73 & 0.74 & 0.83 & 0.46 & 0.52 & 0.45 & 0.59 \\
\end{tabular}

%% file: tables/cross/eqtransformer_phase_test.tex
\setlength{\tabcolsep}{2pt}
\begin{tabular}{c|c|c|c|c|c|c|c|c}
\backslashbox{Data}{Target} & \multicolumn{1}{|c}{ETHZ} & \multicolumn{1}{|c}{INSTANCE} & \multicolumn{1}{|c}{Iquique} & \multicolumn{1}{|c}{SCEDC} & \multicolumn{1}{|c}{STEAD} & \multicolumn{1}{|c}{GEOFON} & \multicolumn{1}{|c}{NEIC} & \multicolumn{1}{|c}{$\diameter$} \\
 & MCC & MCC & MCC & MCC & MCC & MCC & MCC & MCC\\
\hline
ETHZ & 0.97 & 0.96 & 0.98 & 0.88 & 0.99 & 0.67 & 0.73 & 0.88 \\
INSTANCE & 0.95 & 0.97 & 0.98 & 0.87 & 0.99 & 0.72 & 0.78 & 0.89 \\
Iquique & 0.61 & 0.80 & 0.99 & 0.55 & 0.90 & 0.59 & 0.66 & 0.73 \\
SCEDC & 0.96 & 0.95 & 0.97 & 0.96 & 0.99 & 0.68 & 0.72 & 0.89 \\
STEAD & 0.95 & 0.95 & 0.99 & 0.89 & 1.00 & 0.76 & 0.77 & 0.90 \\
GEOFON & 0.58 & 0.58 & 0.89 & 0.28 & 0.58 & 0.82 & 0.61 & 0.62 \\
NEIC & 0.76 & 0.81 & 0.91 & 0.53 & 0.81 & 0.88 & 0.96 & 0.81 \\
\hline
$\diameter$ & 0.83 & 0.86 & 0.96 & 0.71 & 0.89 & 0.73 & 0.75 \\
\end{tabular}

%% file: tables/cross/gpd_phase_test.tex
\setlength{\tabcolsep}{2pt}
\begin{tabular}{c|c|c|c|c|c|c|c|c}
\backslashbox{Data}{Target} & \multicolumn{1}{|c}{ETHZ} & \multicolumn{1}{|c}{INSTANCE} & \multicolumn{1}{|c}{Iquique} & \multicolumn{1}{|c}{SCEDC} & \multicolumn{1}{|c}{STEAD} & \multicolumn{1}{|c}{GEOFON} & \multicolumn{1}{|c}{NEIC} & \multicolumn{1}{|c}{$\diameter$} \\
 & MCC & MCC & MCC & MCC & MCC & MCC & MCC & MCC\\
\hline
ETHZ & 0.88 & 0.81 & 0.92 & 0.27 & 0.85 & 0.27 & 0.62 & 0.66 \\
INSTANCE & 0.88 & 0.90 & 0.95 & 0.43 & 0.88 & 0.34 & 0.68 & 0.72 \\
Iquique & 0.74 & 0.70 & 0.97 & 0.20 & 0.74 & 0.24 & 0.55 & 0.59 \\
SCEDC & 0.57 & 0.47 & 0.69 & 0.88 & 0.56 & 0.09 & 0.19 & 0.49 \\
STEAD & 0.89 & 0.87 & 0.96 & 0.48 & 0.95 & 0.51 & 0.69 & 0.77 \\
GEOFON & 0.38 & 0.53 & 0.68 & 0.17 & 0.47 & 0.66 & 0.58 & 0.50 \\
NEIC & 0.78 & 0.83 & 0.92 & 0.32 & 0.85 & 0.72 & 0.84 & 0.75 \\
\hline
$\diameter$ & 0.73 & 0.73 & 0.87 & 0.39 & 0.76 & 0.40 & 0.59 \\
\end{tabular}

%% file: tables/cross/gpdpick_phase_test.tex
\setlength{\tabcolsep}{2pt}
\begin{tabular}{c|c|c|c|c|c|c|c|c}
\backslashbox{Data}{Target} & \multicolumn{1}{|c}{ETHZ} & \multicolumn{1}{|c}{INSTANCE} & \multicolumn{1}{|c}{Iquique} & \multicolumn{1}{|c}{SCEDC} & \multicolumn{1}{|c}{STEAD} & \multicolumn{1}{|c}{GEOFON} & \multicolumn{1}{|c}{NEIC} & \multicolumn{1}{|c}{$\diameter$} \\
 & MCC & MCC & MCC & MCC & MCC & MCC & MCC & MCC\\
\hline
ETHZ & 0.92 & 0.85 & 0.93 & 0.32 & 0.90 & 0.29 & 0.63 & 0.69 \\
INSTANCE & 0.93 & 0.95 & 0.96 & 0.44 & 0.95 & 0.35 & 0.70 & 0.76 \\
Iquique & 0.76 & 0.77 & 0.98 & 0.23 & 0.79 & 0.35 & 0.58 & 0.64 \\
SCEDC & 0.58 & 0.53 & 0.71 & 0.93 & 0.65 & 0.09 & 0.18 & 0.52 \\
STEAD & 0.91 & 0.91 & 0.97 & 0.51 & 0.99 & 0.50 & 0.69 & 0.78 \\
GEOFON & 0.42 & 0.56 & 0.67 & 0.17 & 0.49 & 0.67 & 0.58 & 0.51 \\
NEIC & 0.79 & 0.85 & 0.95 & 0.23 & 0.87 & 0.74 & 0.84 & 0.75 \\
\hline
$\diameter$ & 0.76 & 0.77 & 0.88 & 0.41 & 0.81 & 0.43 & 0.60 \\
\end{tabular}

%% file: tables/cross/phasenet_phase_test.tex
\setlength{\tabcolsep}{2pt}
\begin{tabular}{c|c|c|c|c|c|c|c|c}
\backslashbox{Data}{Target} & \multicolumn{1}{|c}{ETHZ} & \multicolumn{1}{|c}{INSTANCE} & \multicolumn{1}{|c}{Iquique} & \multicolumn{1}{|c}{SCEDC} & \multicolumn{1}{|c}{STEAD} & \multicolumn{1}{|c}{GEOFON} & \multicolumn{1}{|c}{NEIC} & \multicolumn{1}{|c}{$\diameter$} \\
 & MCC & MCC & MCC & MCC & MCC & MCC & MCC & MCC\\
\hline
ETHZ & 0.91 & 0.89 & 0.93 & 0.62 & 0.95 & 0.33 & 0.59 & 0.75 \\
INSTANCE & 0.90 & 0.94 & 0.96 & 0.54 & 0.98 & 0.49 & 0.68 & 0.78 \\
Iquique & 0.71 & 0.75 & 0.96 & 0.48 & 0.89 & 0.40 & 0.57 & 0.68 \\
SCEDC & 0.76 & 0.76 & 0.82 & 0.91 & 0.84 & 0.13 & 0.29 & 0.64 \\
STEAD & 0.88 & 0.92 & 0.94 & 0.63 & 0.99 & 0.46 & 0.67 & 0.78 \\
GEOFON & 0.40 & 0.42 & 0.66 & 0.17 & 0.40 & 0.51 & 0.45 & 0.43 \\
NEIC & 0.83 & 0.86 & 0.95 & 0.43 & 0.93 & 0.69 & 0.81 & 0.79 \\
\hline
$\diameter$ & 0.77 & 0.79 & 0.89 & 0.54 & 0.85 & 0.43 & 0.58 \\
\end{tabular}

%% file: tables/resampled/detection_test_geofon.tex
\setlength{\tabcolsep}{2pt}
\begin{tabular}{c|c|c|c|c|c|c|c|c}
\backslashbox{Data}{Model} & \multicolumn{1}{|c}{BasicPhaseAE} & \multicolumn{1}{|c}{CRED} & \multicolumn{1}{|c}{DPP} & \multicolumn{1}{|c}{EQTransformer} & \multicolumn{1}{|c}{GPD-Org} & \multicolumn{1}{|c}{GPD} & \multicolumn{1}{|c}{PhaseNet} & \multicolumn{1}{|c}{$\diameter$} \\
 & AUC & AUC & AUC & AUC & AUC & AUC & AUC & AUC\\
\hline
ETHZ & 0.72 & 0.77 & 0.77 & 0.85 & 0.86 & 0.88 & 0.86 & 0.82 \\
INSTANCE & 0.73 & 0.76 & 0.74 & 0.75 & 0.90 & 0.91 & 0.87 & 0.81 \\
Iquique & 0.61 & 0.72 & 0.71 & 0.60 & 0.83 & 0.83 & 0.82 & 0.73 \\
SCEDC & 0.69 & 0.74 & 0.72 & 0.83 & 0.83 & 0.84 & 0.84 & 0.79 \\
STEAD & 0.70 & 0.60 & 0.70 & 0.62 & 0.85 & 0.85 & 0.80 & 0.73 \\
\hline
$\diameter$ & 0.69 & 0.72 & 0.73 & 0.73 & 0.86 & 0.86 & 0.84 \\
\end{tabular}

%% file: tables/resampled/phase_test_geofon.tex
\setlength{\tabcolsep}{2pt}
\begin{tabular}{c|c|c|c|c|c|c|c}
\backslashbox{Data}{Model} & \multicolumn{1}{|c}{BasicPhaseAE} & \multicolumn{1}{|c}{DPP} & \multicolumn{1}{|c}{EQTransformer} & \multicolumn{1}{|c}{GPD-Org} & \multicolumn{1}{|c}{GPD} & \multicolumn{1}{|c}{PhaseNet} & \multicolumn{1}{|c}{$\diameter$} \\
 & MCC & MCC & MCC & MCC & MCC & MCC & MCC\\
\hline
ETHZ & 0.20 & 0.39 & 0.78 & 0.43 & 0.56 & 0.46 & 0.47 \\
INSTANCE & 0.47 & 0.58 & 0.83 & 0.63 & 0.69 & 0.65 & 0.64 \\
Iquique & 0.18 & 0.45 & 0.15 & 0.38 & 0.42 & 0.44 & 0.34 \\
SCEDC & 0.18 & 0.15 & 0.73 & 0.02 & 0.05 & 0.14 & 0.21 \\
STEAD & 0.43 & 0.47 & 0.65 & 0.60 & 0.53 & 0.50 & 0.53 \\
\hline
$\diameter$ & 0.29 & 0.41 & 0.63 & 0.41 & 0.45 & 0.44 \\
\end{tabular}

%% file: tables/resampled/phase_test_neic.tex
\setlength{\tabcolsep}{2pt}
\begin{tabular}{c|c|c|c|c|c|c|c}
\backslashbox{Data}{Model} & \multicolumn{1}{|c}{BasicPhaseAE} & \multicolumn{1}{|c}{DPP} & \multicolumn{1}{|c}{EQTransformer} & \multicolumn{1}{|c}{GPD-Org} & \multicolumn{1}{|c}{GPD} & \multicolumn{1}{|c}{PhaseNet} & \multicolumn{1}{|c}{$\diameter$} \\
 & MCC & MCC & MCC & MCC & MCC & MCC & MCC\\
\hline
ETHZ & 0.50 & 0.62 & 0.71 & 0.63 & 0.68 & 0.60 & 0.62 \\
INSTANCE & 0.55 & 0.68 & 0.74 & 0.72 & 0.75 & 0.68 & 0.69 \\
Iquique & 0.49 & 0.56 & 0.52 & 0.56 & 0.58 & 0.50 & 0.54 \\
SCEDC & 0.31 & 0.33 & 0.67 & 0.12 & 0.14 & 0.26 & 0.30 \\
STEAD & 0.53 & 0.60 & 0.69 & 0.68 & 0.66 & 0.58 & 0.62 \\
\hline
$\diameter$ & 0.48 & 0.56 & 0.67 & 0.54 & 0.56 & 0.52 \\
\end{tabular}